\documentclass[aps,twocolumn,amsmath,amssymb,floatfix,pra,reprint,footinbib,superscriptaddress,showpacs,longbibliography]{revtex4-1}

\usepackage{txfonts}
\usepackage{mathrsfs,times}
\usepackage{ulem}
\usepackage{textcomp}

\newcommand{\sdot}{\,\bullet}
\newcommand{\half}{\frac{1}{2}}
\newcommand{\gentr}{\text{Tr}}
\newcommand{\dd}{d}

\usepackage{graphicx}
\usepackage{epstopdf}
\usepackage[applemac]{inputenc}
\usepackage[T1]{fontenc}
\usepackage[english]{babel}
\usepackage{ae}
\usepackage{siunitx}
\usepackage{color}
\usepackage{url}
\usepackage{amsmath,amssymb,natbib}
\usepackage{psfrag}
\usepackage{fixmath}
\usepackage{booktabs}
\usepackage{slashed}
\usepackage[americaninductors]{circuitikz}
\usepackage{tikz}
\usetikzlibrary{arrows}

\usepackage[colorlinks]{hyperref}
\hypersetup{%
        plainpages=true,
        breaklinks=true,% not default in dvips mode, so we must specify
        hypertexnames=false,%not ideal, but needed when pagenums duplicate (`i' vs. `1')
        pageanchor=true,
        colorlinks=true,
        linkcolor={blue},
        citecolor={red},
        urlcolor={blue},
        anchorcolor={black}
      }

\newcommand{\ket}[1]{\left|#1\right\rangle}
\newcommand{\bra}[1]{\left\langle#1\right|}
\newcommand{\be}{\begin{equation}}
\newcommand{\ee}{\end{equation}}
\newcommand{\bea}{\begin{eqnarray}}
\newcommand{\eea}{\end{eqnarray}}
\newcommand{\Z}{\ensuremath{\mathbb{Z}}}

\newcommand{\ketbra}[2]{\left| #1 \rangle \langle #2 \right|}
\newcommand{\brakket}[3]{\left\langle #1\left| #2 \right| #3\right\rangle}
\newcommand{\expec}[1]{\left\langle #1 \right\rangle}

\newcommand{\abs}[1]{\left|#1\right|}
\newcommand{\abssq}[1]{\left| #1 \right|^2}
\newcommand{\comm}[2]{\left[ #1, #2 \right]}

\newcommand{\sz}{\sigma_z}
\newcommand{\sm}{\sigma_-}
\renewcommand{\sp}{\sigma_+}
\newcommand{\nn}{\nonumber}
\newcommand{\figref}[1]{\mbox{Fig.~\ref{#1}}}

\newcommand{\secref}[1]{\mbox{Section~\ref{#1}}}

\newcommand{\appref}[1]{\mbox{Appendix~\ref{#1}}}
\renewcommand{\eqref}[1]{\mbox{Eq.~(\ref{#1})}}

\newcommand{\alg}[1]{{\color[rgb]{0.6,0,0}#1}}
\makeatletter
\newcommand*{\rom}[1]{\expandafter\@slowromancap\romannumeral #1@}
\makeatother

\begin{document}

\title { The giant acoustic atom -- a single quantum system with a deterministic time delay }

\author{Lingzhen Guo}
\affiliation{Department of Microtechnology and Nanoscience (MC2), Chalmers University of Technology, SE-41296 G\"oteborg, Sweden}
\affiliation{Institut f\"ur Theoretische Festk\"orperphysik, Karlsruhe Institute of Technology (KIT), D-76131 Karlsruhe, Germany}

\author{Arne Grimsmo}
\affiliation{Institut quantique and D\'epartement de Physique, Universit\'e de Sherbrooke, Sherbrooke, Qu\'ebec J1K 2R1, Canada}

\author{Anton \surname{Frisk Kockum}}
\affiliation{Center for Emergent Matter Science, RIKEN, Wako-shi, Saitama 351-0198, Japan}
\affiliation{Department of Microtechnology and Nanoscience (MC2), Chalmers University of Technology, SE-41296 G\"oteborg, Sweden}

\author{Mikhail Pletyukhov}
\affiliation{Institute for Theory of Statistical Physics, RWTH Aachen University, 52056 Aachen, Germany}

\author{G\"oran Johansson}
\affiliation{Department of Microtechnology and Nanoscience (MC2), Chalmers University of Technology, SE-41296 G\"oteborg, Sweden}

\date{\today}

\begin{abstract}
We investigate the quantum dynamics of a single transmon qubit coupled to surface acoustic waves (SAWs) via two distant connection points. Since the acoustic speed is five orders of magnitude slower than the speed of light, the travelling time between the two connection points needs to be taken into account. Therefore, we treat the transmon qubit as a giant atom with a deterministic time delay. We find that the spontaneous emission of the system, formed by the giant atom and the SAWs between its connection points, initially decays polynomially in the form of pulses instead of a continuous exponential decay behaviour, as would be the case for a small atom. We obtain exact analytical results for the scattering properties of the giant atom up to two-phonon processes by using a diagrammatic approach. We find that two peaks appear in the inelastic (incoherent) power spectrum of the giant atom, a phenomenon which does not exist for a small atom. The time delay also gives rise to novel features in the reflectance, transmittance, and second-order correlation functions of the system. Furthermore, we find the short-time dynamics of the giant atom for arbitrary drive strength by a numerically exact method for open quantum systems with a finite-time-delay feedback loop.
\end{abstract}


\maketitle

\section{Introduction}
Superconducting circuits \cite{Blais2004, Devoret2013} form a promising technology for realising the computational nodes of a large-scale quantum network \cite{Kimble2008, Ladd2010}. In a large network, time delays are unavoidable and have to be understood and handled with care. The on-chip delays of standard superconducting microwave circuits, however, are negligible due to the cm chip size and the speed of light. A key element to realize a quantum network is a coherent transducer capable of converting quantum information from the microwave regime to the optical regime, where it could be transmitted over large distances. While many different designs for such a transducer are very actively investigated \cite{Palomaki2013, Andrews2014, Bochmann2013, Hafezi2012, Hisatomi2016, Shumeiko2016}, no high-efficiency solution has been experimentally realized so far. However, there is another possibility to investigate time delays due to propagation on-chip. By transforming the quantum information into surface acoustic waves (SAWs) \cite{Gustafsson2014, Aref2016, Datta1986, Morgan2007, Manenti2017}, \textit{i.e.}, sound waves travelling with a velocity five orders of magnitude lower than the speed of light, microsecond propagation-time delays can easily be achieved. Delay lines is indeed also one of the main applications of SAWs in microwave technology \cite{Morgan2007}.

The theoretical description of open quantum systems including propagation time delays has until recently focussed on so-called cascaded quantum systems \cite{Gardiner1993, Carmichael1993, Gardiner2004, Gough2009, Gough2009a}, where no closed loops for quantum information are created. There are also tools to describe systems with closed loops, but where the dynamics of quantum systems is approximatively coherent during the short time-delay, which can then be included in terms of signal phase shifts \cite{Gough2009, Gough2009a, Kerckhoff2012, Kockum2014}. Recently, there has been an increased interest in longer time delays and closed loops, where the quantum systems have time to evolve dissipatively and really emit energy into the transmission line before that energy returns \cite{Dorner2002, Tufarelli2013, Laakso2014, Grimsmo2015, Fang2015, Pichler2016, Ramos2016}.

In this paper, we analyze one of the simplest examples of an open quantum system with a deterministic
propagation-time delay. It consists of a single two-level atom, connected at two points to a single open transmission line. This is not only conceptually one of the simplest examples but also straightforward to implement experimentally \cite{Gustafsson2014, Aref2016, Manenti2017}. The single two-level quantum system is realized by a transmon qubit \cite{Koch2007} with strong nonlinearity, which serves as an artificial atom. Compared to the SAW wavelength ($\lesssim$\SI{1}{\textmu\meter}), the transmon can be designed to extend over distances at least several hundred, or even a thousand, times greater($>$\SI{100}{\textmu\meter}). Since the speed of SAWs is five orders of magnitude slower than the speed of light, we need to consider the time delay between the two connection points. However, we neglect the time delay in the connections themselves and the transmon, since they are metallic and signals travel through them at the speed of light. Therefore, we refer to our single two-level quantum system as a \textit{giant atom}. Other examples of simple systems that can exhibit time delay are a two-level atom in front of a mirror \cite{Dorner2002, Tufarelli2013, Pichler2016,Scarani2016} and two two-level atoms at a long distance from each other in an open transmission line \cite{Milonni1974, Rist2008, VanLoo2013, Laakso2014, Fang2014, Fang2015, Pichler2016}.

The two connection points of the giant atom introduce boundary conditions for the SAWs propagating in the open transmission line, making the area between the connection points reminiscent of a cavity. However, it is different from both the cavities used for optical photons in cavity quantum electrodynamics (QED) and the transmission line resonators used for microwave photons in circuit QED, as well as from the effective cavities for single photons that can be formed by several small atoms in an open transmission line \cite{Rist2008, Chang2012, Fratini2014}. For example, we find that the total energy stored in atom and the SAWs between the connection points exhibits an intially polynomial decay process before reverting to exponential decay in the long-time limit.

\section{The Model}
\label{sec:TheModel}

\begin{figure}
\centering
\includegraphics[width=\linewidth]{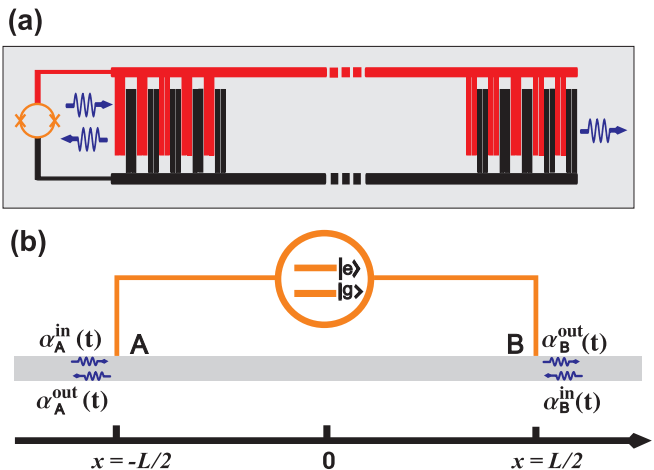}
\caption{{\bf Giant atom coupled to SAWs.}
(a) The giant atom is a transmon (orange SQUID circle) with large interdigitated capacitance (red and black structures) from its two islands. The giant atom couples to SAWs propagating on the piezoelectric substrate (grey color) through the piezoelectric effect.
(b) The transmon qubit is modelled by a two-level atom with two legs (labelled by $A$ and $B$, respectively) coupled to a one-dimensional (1D) transmission line. The distance between the two legs is $L$. In- and out-going phonons, including external drive fields, near the two connection points are shown as blue arrows. \label{fig:Device}}
\end{figure}

In \figref{fig:Device}(a), we show the setup investigated in this work. A transmon is coupled to a piezoelectric substrate (grey color). The interdigitated capacitance forming the two islands of the transmon also forms a transducer which couples to the SAWs propagating on the substrate. In our case, the whole interdigital transducer (IDT) consists of two local IDTs at its two ends, far away from each other. Each IDT has $N$ pairs of fingers ($N=5$ pairs are shown in \figref{fig:Device}(a)). The fork configuration of each finger is designed to minimize internal mechanical reflections \cite{Gustafsson2014, Aref2016, Datta1986, Morgan2007}.

In this paper, we will explore the transmon dynamics in the qubit regime, \textit{i.e.}, where only the lowest two transmon levels with energy splitting $\hbar \omega_0$ are involved. Considering just one of the two local IDTs, we assume the relaxation rate corresponding to a single IDT finger pair is $\gamma_0$. Due to the interference of all the finger pairs, the total effective relaxation rate of each local IDT is given by \cite{Kockum2014}
\be
\gamma = \gamma_0 \frac{1-\cos \left(N \omega_0 \tau \right)}{1 - \cos \left(\omega_0 \tau \right)},
\label{gamma}
\ee
where $\tau$ is the travelling time (negligible in this case) between neighbouring finger pairs (of the same colour shown in \figref{fig:Device}(a)). We now consider the distance between neighbouring finger pairs to match the wavelength of the corresponding phonons, so that the transmon-phonon coupling is maximized, {\it i.e.}, $\omega_0\tau = 2n\pi, \ n\in\Z^+$, leading to $\gamma = N^2 \gamma_0$. The bandwidth of phonons involved in the qubit dynamics will be determined by the coupling $\gamma$.  In the following, we will consider the regime $\gamma \ll \omega_0/N$, so that we can neglect the weak frequency dependence of the coupling around the maximum in \eqref{gamma}.

The distance $L$ between the centers of the two local IDTs can be made long, straightforwardly up to a few thousands of SAW wavelengths \cite{Aref2016}, which is the regime of interest in this paper. We characterize this distance by the corresponding time delay $T=L/v_g$ for the SAWs travelling with the velocity $v_g$ on the piezoelectric substrate.We thus arrive at the model of a two-legged giant atom, as sketched in \figref{fig:Device}(b), with two legs labelled by $A$ and $B$, respectively. The model Hamiltonian is
\bea
&&\mathscr{H}=\hbar\omega_0 \ketbra{e}{e} + \sum_{\alpha=1,2} \int \hbar \omega_p a^{\dagger}_{\alpha \omega_p} a_{\alpha \omega_p} d\omega_p \nonumber\\
&&+ \sum_{\alpha=1,2}\int \hbar\sqrt{\frac{\gamma}{4\pi}}\Big[\sigma_-a^{\dagger}_{\alpha \omega_p}\Big(e^{-ic_\alpha k_p \frac{L}{2}}+e^{ic_\alpha k_p\frac{ L}{2}}\Big)+\mathrm{h.c.}\Big]d\omega_p, \quad
\label{Hstart}
\eea
where $a_{\alpha\omega_p}$ are phonon field operators for the right- ($\alpha=1$) and left-propagating ($\alpha=2$) phonons satisfying $ \comm{a_{\alpha\omega_p}}{a^\dag_{\alpha'\omega'_p}} = \delta_{\alpha\alpha'}\delta(\omega_p-\omega'_p)$, and $\omega_p$ represents the frequency of phonon modes. We have defined atomic operators $\sm \equiv \ketbra{g}{e}$ and $\sp \equiv (\sm)^\dagger$ where $\ket{g}$ and $\ket{e}$ are the atomic ground and excited states. The coupling term in the second line of \eqref{Hstart} has included the phase difference between two legs of giant atom at positions $x = - L/2$ and $x = L/2$. The parameter $k_p$ is the wave vector of the SAWs, \textit{i.e.}, $k_p=\omega_p/v_g$ or $k_p=2\pi /\lambda_{SAW}$ with $\lambda_{SAW}$ being the wavelength of the SAWs. Since the typical value of $\gamma$ is tens of MHz, which is small compared to $\omega_0$ (several GHz), it is reasonable to assume that the SAW dispersion is flat over the atom's bandwidth. The notation $c_{\alpha=1} =+1$, $c_{\alpha=2} =-1$ is used to distinguish the interaction of the giant atom with the right- and left-propagating phonon fields, respectively.

In \secref{sec:SinglePhononProcesses}, we will explore the spontaneous-emission dynamics of the giant atom when it is excited directly by an electrical gate \cite{Gustafsson2014}, and also the single-phonon scattering properties of the giant atom when it is driven by SAWs emanating from external IDTs. We then extend the scattering calculations to two-phonon processes in \secref{sec:TwoPhononProcesses}, allowing us to study second-order correlation functions for the scattered phonons. Finally, in \secref{sect:cascade}, we investigate the short-time dynamics of the giant atom when it is subjected to coherent driving of arbitrary strength.

\section{Single-phonon processes}
\label{sec:SinglePhononProcesses}

We will first consider the single-excitation subspace of the giant atom's dynamics, since this is amenable to analytic solutions. In this subspace, the total state of the two-level giant atom and SAW field in the transmission line can be described by \cite{Peropadre2011}
\be
\ket{\Psi(t)} = \int d\omega \left[ \alpha_{1\omega}(t) a^\dag_{1 \omega} + \alpha_{2 \omega}(t) a^\dag_{2\omega} \right] \ket{g, vac} + e(t) \ket{e, vac},
\label{Psit}
\ee
where $\ket{vac}$ represents the ground state of SAW field in the transmission line. The integral part describes the state of a single phonon propagating in the transmission line towards the right, $\alpha_{1\omega}$, or the left, $\alpha_{2\omega}$, with the giant atom in the ground state $\ket{g}$. When the phonon is absorbed, the giant atom is in the excited state $\ket{e}$ with the probability amplitude $e(t)$. From the Schr\"odinger equation $i\hbar\partial/\partial t|\Psi(t)\rangle=\mathscr{H}|\Psi(t)\rangle$, we obtain the evolution of $e(t)$ (see \appref{app:SinglePhononEOM} for details)
\bea
\frac{\partial e(t)}{\partial t} &=& -i \omega_0 e(t) - \gamma \left[ e(t) + e (t-T) \right] \nn\\
&& -i V \left[ \alpha^{in}_A(t) + \alpha^{in}_A(t-T) + \alpha^{in}_B(t) + \alpha^{in}_B(t-T) \right].
\label{EOMb}
\eea
The first term in the right-hand side (RHS) of \eqref{EOMb} describes the unitary evolution of the giant atom without dissipation and driving. The second term in the RHS of \eqref{EOMb} describes the relaxation process via the two legs of the giant atom, which includes the time delay between the two legs. The last term in the RHS of \eqref{EOMb} describes the dynamics due to external driving sources exciting the atom through both leg $A$ and leg $B$. If a plane-wave driving field only comes from the leg $A$, as shown in \figref{fig:Device}(a), the driving terms in \eqref{EOMb} is $\alpha^{in}_A(t) = A e^{-i \omega_d t}$ and $\alpha^{in}_B(t) = 0$ with $A$ the amplitude of the drive. The coherent coupling amplitude $V$ is determined by the relaxation rate $\gamma$ through $V=\sqrt{\gamma v_g/2}$ (details are given in \appref{app:SinglePhononEOM}). Previous work on the giant atom only considered the Markov limit, where the time delay $T$ is negligible \cite{Kockum2014}. In this article, on the other hand,  we are mainly interested in understanding the effect of a nonnegligible $T$. Below, we specify the parameter regimes that will be considered.

\subsection{Spontaneous Emission}

\subsubsection{Overview of parameter regimes}\label{parameterregimes}

\begin{figure}
\centering
\includegraphics[width=\linewidth]{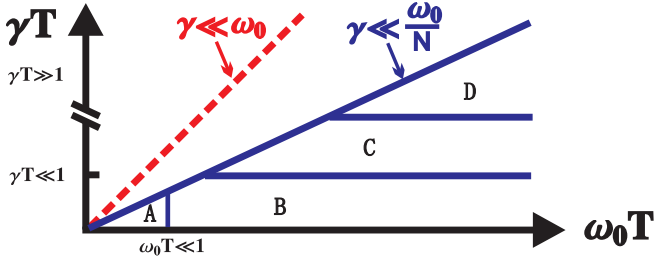}
\caption{{\bf Parameter space}. The parameter region under the line $\gamma\ll\omega_0/N$, where the RWA holds, is divided into several different subregions. Region $A$: The limit of both long wavelength, $\omega_0 T = k L \ll 1$, and Markovian dynamics, $\gamma T \ll 1$. Region $B$: Still Markovian dynamics, but arbitrary phase difference $\omega_0 T$. Region $C$: The moderately non-Markovian regime $\gamma T \sim 1$. Region $D$: The deep non-Markovian regime, $\gamma T \gg 1$.
\label{fig:ParameterRegimes}}
\end{figure}

Equation (\ref{EOMb}) is a time-delay differential equation. Without the external driving, \textit{i.e.}, setting $\alpha^{in}_{A}(t)=0$ and $\alpha^{in}_{B}(t)=0$, we straightforwardly find the analytical solution describing the spontaneous relaxation of the giant atom's excitation amplitude (see \appref{app:SinglePhononSpontaneousEmission}, and also Ref.~\cite{Dorner2002}, for details)
\be
e(t) = e(0) e^{-i(\omega_0 T - i \gamma T) \frac{t}{T}} \sum_{n = 0}^{[t/T]} \frac{(\gamma T)^n}{n!} \left(n - \frac{t}{T}\right)^n e^{i n (\omega_0 T - i \gamma T)},
\label{et}
\ee
where $e(0)$ is the initial probability amplitude of giant atom and $[t/T]$ is the integer part of $t/T$. From \eqref{et}, we see that $e(t)$ is a function of the dimensionless time ${t}/{T}$, depending on the two dimensionless parameters $\gamma T$ and $\omega_0 T$. Therefore, the relaxation properties of the giant atom is determined by the parameter plane spanned by $\gamma T$ and $\omega_0 T$, which can be divided into several regions as shown in \figref{fig:ParameterRegimes}. Since we work in the rotating wave approximation (RWA), we only consider the region under the line $\gamma \ll \omega_0$. Furthermore, we neglect the frequency dependence of the local IDTs (see \eqref{gamma}), implying that $\gamma \ll \omega_0/N$. We divide the remaining region into several different subregions marked by $A$, $B$, $C$ and $D$, respectively.

The corner region $A$ is defined by the condition $\omega_0 T\ll 1$. Since $\omega_0 T=k_pL =2\pi L/\lambda_{SAW}$, the condition $\omega_0 T \ll 1$ corresponds to the long wavelength limit $\lambda_{SAW} \gg 2\pi L$, which means we can neglect the phase acquired by SAWs travelling between the connection points. Due to the RWA condition $\gamma \ll \omega_0$, this region is also in the Markov limit, \textit{i.e.}, $\gamma T \ll \omega_0 T \ll 1$. Thus, we can neglect high orders of $\gamma T$ ($n \geq 1$) in \eqref{et} and obtain an approximate result, $e(t) \approx e(0) \exp(-i \omega_0 t - 2 \gamma t)$. This is indeed the result for a small atom with total relaxation rate $2\gamma$, since the inter-leg distance is much smaller than the phonon wavelength. Here, we also note that this regime cannot be accessed experimentally in the SAW-transmon system, since each individual local IDT already consists of $N$ legs separated by the phonon wavelength. However, since the wavelength of microwave photons is usually much longer than the dimensions of a superconducting qubit, most circuit-QED setups work inside this regime.

Parameter region $B$ is also in the Markov limit $\gamma T\ll 1$, but with an arbitrary $\omega_0 T$, which means the phase acquired by SAWs travelling in the transmission line between the two legs needs to be considered \cite{Kockum2014}. In this case, the main contribution comes from the low orders in the series of \eqref{et}. Therefore, by taking $n-t/T \approx -t/T$ in the limit $t\gg T$, we have the asymptotic behaviour $e(t) \approx e(0)\exp(-i \bar{\omega}_0 t - \bar{\gamma} t)$, where the effective frequency is $\bar{\omega}_0 = \omega_0 + \gamma e^{\gamma T} \sin(\omega_0 T)$ and the effective decay rate is given by $\bar{\gamma} = \gamma[1 + e^{\gamma T} \cos(\omega_0 T)]$. Considering the Markov limit $\gamma T\ll 1$, we can further take $e^{\gamma T}\approx1$ in $\bar{\omega}_0$ and $\bar{\gamma}$. We see that $\bar{\omega}_0$ and $\bar{\gamma}$ are both modified by the phase factor $\omega_0 T$ and the results coincide with those given in Ref.~\cite{Kockum2014}, where the distances between the legs were represented by frequency-dependent phase shifts.

\begin{figure*}
\centering
\includegraphics[width=\linewidth]{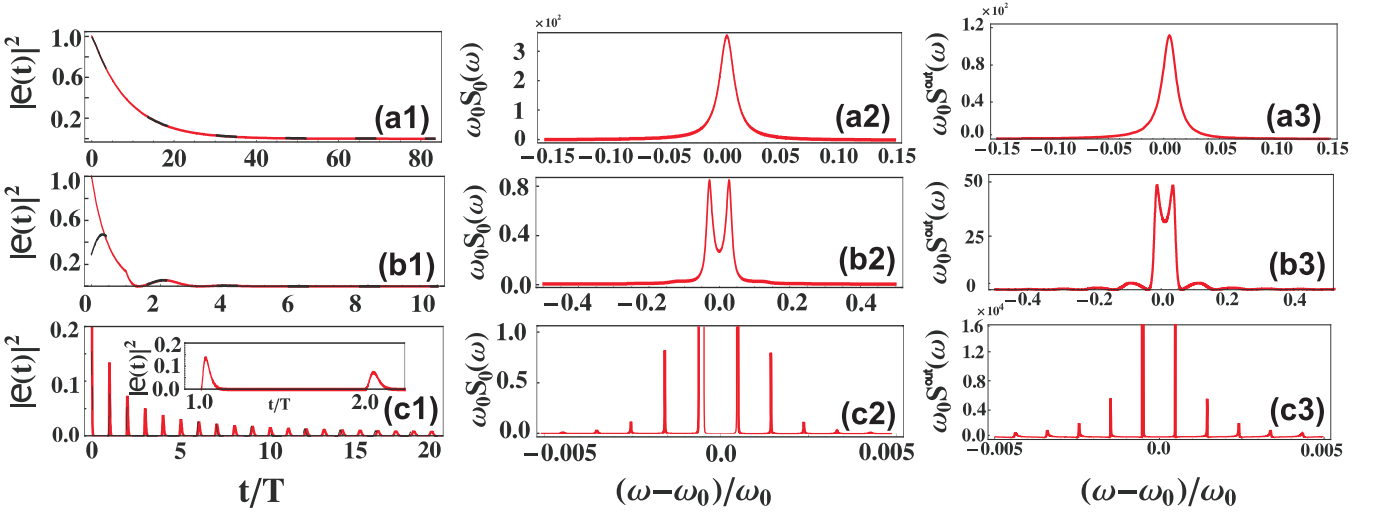}
\caption{\textbf{ Time evolutions and power spectra of giant atom}.
(a1) Time evolution of $\abssq{e(t)}$ in parameter region $B$ with $\gamma T = 0.045, \omega_0 T = 2.4 \pi$. The red line is a numerical simulation and the black dashed line is the analytical result from \eqref{etLaplace} with a single mode $\omega_{(0)}$.
(a2), (a3) The corresponding power spectra of the giant atom and the outgoing phonons, respectively.
(b1) Time evolution of $\abssq{e(t)}$ in parameter region $C$ with $\gamma T = 1.0, \omega_0 T = 20 \pi$. The black dashed line is the analytical result from \eqref{etLaplace} including the two modes $\omega_{(0)}$ and $\omega_{(-1)}$.
(b2), (b3) The corresponding power spectra of the giant atom and the outgoing phonons, respectively.
(c1) Time evolution of $\abssq{e(t)}$ in parameter region $D$ with $\gamma T = 37.5, \omega_0 T = 2000 \pi$. The black dashed line is the analytical result from \eqref{etLaplace} including eleven modes, \textit{i.e.}, from $\omega_{(-10)}$ to $\omega_{(10)}$. The evolution exhibits revival peaks, the first two of which are shown in more detail in the inset.
(c2), (c3) The corresponding power spectra of the giant atom and the outgoing phonons, respectively.
\label{figSpectrum}}
\end{figure*}

The parameter regions $C$ and $D$ are both beyond the Markov approximation. Region $C$ corresponds to a moderate non-Markovian regime $\gamma T \sim 1$ while region $D$ is the deep non-Markovian regime $\gamma T\gg 1$. In region $D$, the dominant term is the highest order in the series of \eqref{et}. Thus we have the approximate solution in the time interval $mT \leq t < (m+1) T$, with $m\in\Z^+$,
\be
e_m(t) \approx e(0) e^{-i(\omega_0 T - i \gamma T) \frac{t}{T}} \frac{(\gamma T)^m}{m!} \left(m - \frac{t}{T} \right)^m e^{i m (\omega_0 T - i \gamma T)}.
\label{bm}
\ee
In region $C$, no such simplifications are possible and the complete time evolution, given by Eq.(\ref{et}), must be used. Examples of relaxation dynamics in the experimentally accessible parameter regions $B$, $C$ and $D$ are given in Figs.~\ref{figSpectrum}(a1), (b1) and (c1), respectively, which we will discuss in detail below. However, to understand the dynamics better, it is useful to first also study the power spectrum of the giant atom.

\subsubsection{Power spectra}

We now study the solution of \eqref{EOMb} from another point of view by decomposing $e(t)$ into a superposition of many independent modes, \textit{i.e.}, $e(t) = \sum_k c_k e^{-i \omega_{(k)} t}$. In general, the mode frequencies $\omega_{(k)}$ can be complex numbers where the imaginary part gives the relaxation rate of each mode. Plugging this form into \eqref{EOMb} without driving terms, we obtain the analytical solution
\be
\omega_{(k)} = \omega_0 - i \gamma + i \frac{1}{T} W_k \left(-\gamma T e^{\gamma T + i \omega_0 T}\right),
\label{LambertW}
\ee
with $k\in \Z$. Here, $W(z)$ is the Lambert W-function \cite{Corless1996} defined by the equation $z=W(z)e^{W(z)}$, which in general is a multivalued function with branches $W_k(z)$, $k\in \Z$. By Fourier transforming \eqref{EOMb}, we obtain the solution of $e(t)$ for $t>0$ (details are given in \appref{app:SinglePhononSpontaneousEmission})
\be
e(t) = e(0) \sum_k \frac{e^{-i \omega_{(k)} t}}{1 - \gamma T e^{i \omega_{(k)} T}} .
\label{etLaplace}
\ee
We assume the giant atom is initially in the excited state and set $e(0)=1$ in the following. We now define the power spectrum of the giant atom by
\bea
S_0(\omega) &\equiv& \omega_0 \abssq{\int_{-\infty}^{+\infty} e(t) e^{i\omega t} dt}.
\label{Satom}
\eea
According to the Wiener-Khinchin theorem \cite{Carmichael2008,Scully1997,Scott2012}, the power spectrum can be obtained by Fourier transform of the autocorrelation function $C_{ee^*}(\tau)\equiv\int e^*(t)e(t+\tau) dt$, i.e., $S_0(\omega)= \omega_0\int C_{ee^*}(\tau) e^{i\omega\tau}d\tau$ \cite{Auffeves2008}. From Parseval's theorem, we have the identity $(2\pi)^{-1}\int S_0(\omega) d\omega =\omega_0\int |e(t)|^2dt$, which is a
reflection of energy conservation, i.e., the energy in the time domain is equal to the energy in the frequency domain. Therefore, the power spectrum $S_0(\omega)$ (normalized by a factor) is the density of the atom's energy distribution over the frequency domain. 

From \eqref{etLaplace}, we can calculate the atomic power spectrum as following
\bea
S_0(\omega) &=& \omega_0 \abssq{\sum_k \frac{i }{1 - \gamma T e^{i \omega_{(k)} T}} \frac{1}{\omega - \omega_{(k)}}} \nn\\
&=& \omega_0\frac{1}{\Big|\omega - \omega_0 + i \gamma(1 + e^{i \omega T})\Big|^2}.
\label{Spectrum}
\eea
Here, we have used \eqref{ID} in \appref{app:SinglePhononSpontaneousEmission} to arrive at the second expression. In \figref{figSpectrum}, we show the time evolution of $\abssq{e(t)}$ in different parameter regions and the corresponding power spectra. The red curve in \figref{figSpectrum}(a1) shows the time evolution of $\abssq{e(t)}$ from a numerical simulation with parameters in region $B$. The time evolution can be well fit by an exponential decay (black dashed curve) obtained from \eqref{etLaplace} including only one frequency mode $\omega_{(0)}$. The corresponding atomic power spectrum in \figref{figSpectrum}(a2) shows a single peak. The red curve in \figref{figSpectrum}(b1) shows the time evolution of $\abssq{e(t)}$ with parameters in region $C$. In the atomic power spectrum shown in \figref{figSpectrum}(b2), we see that there are two dominant modes, corresponding to $\omega_{(-1)}$ and $\omega_{(0)}$. The long-time behaviour of $\abssq{e(t)}$ can be fit well by \eqref{etLaplace} including the two modes $\omega_{(-1)}$ and $\omega_{(0)}$ (black dashed curve in \figref{figSpectrum}(b1)). To fit the short-time dynamics, we need to include more modes.

In \figref{figSpectrum}(c1), the red curves show the time evolution of $\abssq{e(t)}$ with parameters in region $D$. We see that the giant atom exhibits revivals during the evolution in intervals spaced by $T$. Here, the distance between the two IDTs is one thousand SAW wavelengths, \textit{i.e.}, $L=1000\lambda_{SAW}$. The giant atom decays to the ground state between revivals since the travelling time between the IDTs is much larger than the decay time of the giant atom ($ T \gg 1/\gamma $). The corresponding atomic power spectrum, shown in \figref{figSpectrum}(c2), has a more complicated structure of multiple peaks with narrow widths. The higher-frequency modes correspond to shorter wavelengths. In \appref{app:SinglePhononSpontaneousEmission}, we derive the approximate solution for complex mode $\omega_{(k)}= \mathrm{Re}[\omega_{(k)}]+i\mathrm{Im}[\omega_{(k)}]$ in parameter region $D$:
\bea
\mathrm{Re}[\omega_{(k)}] &\approx& \omega_0 - \frac{1}{T}\left[ \pi (2k + 1 + \Delta) - \arctan\frac{\pi (2k + 1 + \Delta)}{\gamma T} \right] \nn\\
\mathrm{Im}[\omega_{(k)}] &\approx& -\frac{1}{2T}\ln \left(1 + \left[\frac{(2k + 1 + \Delta)\pi}{\gamma T} \right]^2 \right).
\label{modi.freq.CD}
\eea
Here, the residual phase is defined by $\Delta = \omega_0 T / \pi - 2 n \in [0,2)$ with $n \in \Z$. The real part of $k$th mode $\mathrm{Re}[\omega_{(k)}]$ gives the position of the center of the corresponding peak in the power spectrum. The positions of the peaks are roughly equally spaced; the frequency spacing is given by $\Delta \omega \approx {2\pi}/{T}$ when $\gamma T \gg (2k + 1 + \Delta)\pi$. The imaginary part of $k$th mode $\mathrm{Im}[\omega_{(k)}]$ gives the width of each peak in the power spectrum. In the limit $\gamma T \gg (2k + 1 + \Delta) \pi$, the width of each peak is approximately given by ${\pi^2 (2k + 1 + \Delta)^2}/{\gamma^2 T^3}$.

In an experiment, it is more convenient to measure the power spectrum of the outgoing phonons $\alpha^{out}(t) \equiv\alpha_{A}^{out}(t) = \alpha_{B}^{out}(t)$, which is given by the interference of the phonons emitted from the two legs
\be
\alpha^{out}(t) = - i \frac{V}{v_g} \left[e(t) + e(t-T) \right].
\label{alphaoutt}
\ee
Similar to the definition of the atomic power spectrum in \eqref{Satom}, we define the power spectrum of $\alpha^{out}(t)$ (fluorescence spectrum) and calculate it as (see \appref{app:SinglePhononSpontaneousEmission} and Ref.~\cite{Peropadre2011} for details)
\bea
S^{out}(\omega) &\equiv& v_g \abssq{\int_{-\infty}^{+\infty} \alpha^{out}(t) e^{i \omega t} dt} \nn\\
&=& v_g \abssq{\frac{\gamma}{2V} \sum_k \frac{1}{1 - \gamma T e^{i\omega_{(k)} T}} \frac{1 + e^{i\omega T}}{\omega - \omega_{(k)}}}\nn\\
 &=& \gamma\frac{1 + \cos \omega T}{\abssq{\omega - \omega_0 + i \gamma (1+ e^{i \omega T})}}.
\label{OutputSpectrum}
\eea
Compared to \eqref{Spectrum}, there is an additional factor $(1 + \cos \omega T)$, which comes from the interference of outgoing photons from the two legs of the giant atom. In Figs.~\ref{figSpectrum}(a3), (b3), and (c3), we calculate the power spectra of the outgoing phonons in the three parameter regimes $B$, $C$, and $D$, respectively. The fluorescence spectrum $S^{out}(\omega)$ can be directly obtained by measuring the autocorrelation function of outgoing phonon fields $C_{\alpha\alpha^*}(\tau)\equiv\int\alpha^{out*}(t)\alpha^{out}(t+\tau)dt$. According to the Wiener-Khinchin theorem, the quantity $S^{out}(\omega)$ is given by the Fourier transform of the above autocorrelation function, i.e., $S^{out}(\omega)=v_g \int C_{\alpha\alpha^*}(\tau)e^{i\omega\tau}d\tau$ \cite{Auffeves2008}. The fluorescence spectrum $S^{out}(\omega)$ is the energy distribution of outgoing phonons over the frequency domain.

In the above discussion, we assumed a general case where the parameters were chosen to satisfy $\omega_0 T \neq (2n + 1)\pi$, such that all the modes die out in the end and the giant atom decays to the ground state. In the case of $\omega_0 T = (2n + 1)\pi$, from the definition of the Lambert W-function and \eqref{LambertW}, we have a real frequency mode $\omega_{(0)} = \omega_0$ with zero imaginary part. In the long-time limit, this mode survives while all the other modes die out. Therefore, we have the stationary solution
$
e(t\rightarrow+\infty) = \frac{1}{1 + \gamma T} e^{-i\omega_0 t}.
$
This corresponds to a \textit{dark state} which does not decay to the ground state in spite of the coupling to the open transmission line. The reason is that the emissions from the two legs cancel each other due to the phase difference $\omega_0 T = (2n + 1)\pi$, reminiscent of how two or more atoms in waveguides can form dark states \cite{Lehmberg1970a, Stannigel2012, Lalumiere2013, Pichler2015a}. Therefore, the stationary values of $\abssq{e(+\infty)}$ will be finite and the corresponding spectra $\abssq{S_0(\omega)}$ will show a singularity at $\omega = \omega_0$ due to the zero imaginary part of the dark mode.

\subsubsection{Polynomial decay}

\begin{figure}
\centering
\includegraphics[width=\linewidth]{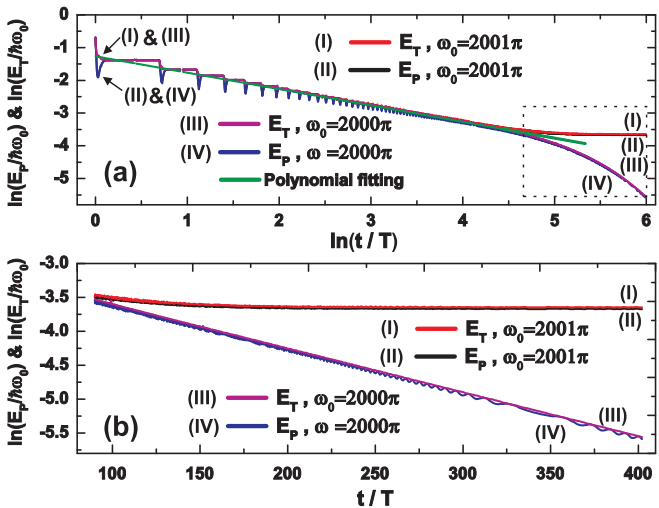}
\caption{\textbf{Spontaneous decay of giant atom.}
(a) The initial decay of the phonon energy stored between the connection points of the giant atom $E_P(t)$ [curves (\rom{2}) and (\rom{4})] and the total energy of giant atom $E_T(t)$ [curves (\rom{1}) and (\rom{3})] exhibit universal polynomial behaviours on the time scale of $T$. However, the long-time behaviour (the part indicated by the dashed box) deviates from this polynomial behaviour.
(b) The long-time decay follows exponential decay laws. The slope (decay exponent) for $\omega_0 T = 2000 \pi$ (bright state) is given by $\pi^2/(\gamma^2 T^2)$. The slope for $\omega_0 T = 2001 \pi$ (dark state) is zero.
\label{figPolynomialFitting}}
\end{figure}

As discussed in the introduction, the whole structure shown in \figref{fig:Device}(a) is reminiscent of a cavity for SAW phonons, since the two connection points of the giant atom introduce boundary conditions that can act as semitransparent mirrors. The energy stored in the atom and in the SAWs between the connection points is gradually lost to SAWs propagating out from the two connection points. In the case of an ordinary cavity with a small atom inside, and in the case of a small atom coupled to an open transmission line, the damping follows an exponential decay law. In our setup, however, we here show that the decay process in parameter region $D$ exhibits a different behaviour.

In \figref{figPolynomialFitting}, we plot the numerical results for the time evolution of the energy stored in the form of phonons between the connection points, $E_P(t)$, and the total energy of the giant atom, $E_T(t)\equiv E_P(t)+\hbar\omega_0|e(t)|^2$, for parameters $\omega_0 T = 2001 \pi$ (dark state) and $\omega_0 T = 2000 \pi$ (bright state). In the long-time limit, the decaying behaviour depends on the parameter $\omega_0T$ [see the part in the dashed box in \figref{figPolynomialFitting}(a)]. However, before this stage, there is a universal energy damping following a polynomial law. From \figref{figPolynomialFitting}(a), we see that the total energy $E_T(t)$ [curves (\rom{1}) and (\rom{3})] decays monotonically in a staircase-like function. The decay of the phonon energy $E_P(t)$ [curves (\rom{2}) and (\rom{4})] overlaps with $E_T(t)$ most of the time except around integer multiples of $T$, where a dip of $E_P(t)$ appears on the time scale of $\gamma^{-1}\ll T$. On the large time scale $T$, we find that the energy decay of the stairs shown in $E_P(t)$ and $E_T(t)$ obeys (see \appref{app:SinglePhononPolynomialDecay} for details)
\be
E_{P/T} \approx \frac{\hbar \omega_0}{2 \sqrt{\pi}} \left(\frac{t}{T}\right)^{-1/2}.
\label{polynomialDecay}
\ee
It is interesting to note that the timescale for the polynomial decay, which here follows an inverse square-root law, is not set by the coupling strength $\gamma$ but is instead determined solely by the propagation time $T$. In contrast, the timescale of the exponential decay of a small atom in an open transmission line would be completely fixed by $\gamma$.

The spontaneous emission of a small atom is a continuous process. However, the staircase behavior of the total energy $E_T(t)$ indicates that the giant atom emits energy in the form of phonon pulses with time period $T$. In an experiment, it should be straightforward to measure the outgoing phonons from the two legs of the giant atom, $\alpha^{out}(t)$. One can measure the total energy of each outgoing phonon pulse, i.e., $E^{out}(m)\equiv v_g\int_{mT}^{(m+1)T}|\alpha^{out}(t)|^2dt$ for $m\leq t/T < m+1$. We find that the total pulse energy $E^{out}(m)$ as a function of the pulse number $m$ also decays polynomially, i.e., $E^{out}(m)/\hbar\omega_0\approx \frac{1}{8\sqrt{\pi}} m^{-3/2}$ [see Eq.~(\ref{Eoutpolynomial}) in \appref{app:SinglePhononPolynomialDecay}]. Beside the polynomial behaviours of $E_P(t)$, $E_T(t)$ and $E^{out}(m)$, we find that the revival peaks shown in Fig.~\ref{figSpectrum}(c1) also exhibit polynomial decay. In \appref{app:SinglePhononPolynomialDecay}, we calculate the time position of the revival peaks' maxima $t_m=mT+m/\gamma $ for $t\in [mT, (m+1)T)$ and the values of the revival peaks $P^{max}_e(t_m)$. As given by Eq.~(\ref{polynomialDecaype}) in \appref{app:SinglePhononPolynomialDecay}, the decay of revival peaks follows a polynomial law $P^{max}_e(t_m)\propto t^{-1}_m$. From Fig.~\ref{figSpectrum}(c1), we also see that the width of the revival peak becomes broader and broader. In fact, from the revival peaks' maxima $t_m=mT+m/\gamma $, we see that the peak will spread over the whole time interval $[mT,(m+1)T]$ when $m>\gamma T$, which implies that the polynomial decay is valid only for $t<\gamma T^2$.  

For sufficiently long time $t>\gamma T^2$, the decaying behaviour deviates from the polynomial law. In parameter region $D$, the total system energy is mainly stored in the form of propagating phonon wave packets excited at both legs of the giant atom as shown in \figref{figSpectrum}(c1). A wave packet contains many frequency modes, given by \eqref{modi.freq.CD}. Each mode $k$ decays at a different rate as described by its imaginary part $\mathrm{Im}[\omega_{(k)}]$. The initial polynomial decaying behaviour is the collective effect of multiple modes decaying. In the long-time limit, however, only the mode with the slowest (exponential) decay rate survives. In \figref{figPolynomialFitting}(b), we show the long-time decaying behaviour in two cases, both following an exponential decay law. From the analytical expression of $\omega_{(k)}$ given by \eqref{modi.freq.CD}, we see the slope for $\omega_0 T = 2000 \pi$ is $\pi^2/(\gamma^2 T^2)$. For the dark state case $\omega_0 T = 2001 \pi$, the smallest imaginary part is zero, explaining the finite remaining energy and zero decay rate.

\subsection{Single-phonon scattering}
\label{sec:SinglePhononScattering}

Above, we studied the spontaneous emission of the undriven giant atom. Now, we investigate the scattering process in the weak-driving limit, \textit{i.e.}, single-phonon driving. In experiments, it is convenient to measure transmittance and reflectance of such a weak drive. We consider SAWs incoming towards leg $A$ and transmitted to leg $B$, \textit{i.e.}, the driving terms in \eqref{EOMb} are set to be $\alpha^{in}_A(t) = A e^{-i\omega_dt}$ and $\alpha^{in}_B(t) = 0$ (see also \figref{fig:Device}(a)). Using the method of Laplace transformation, the reflectance in the long-time limit, defined as $\mathcal{R} = \abssq{\alpha^{out}_A(\infty)}/\abssq{\alpha^{in}_A(t)}$, is calculated to be (see \appref{app:SinglePhononRTCoefficients} for details)
\be
\mathcal{R} = \frac{\gamma^2 \left(1 + \cos \omega_d T \right)^2}{\left[ (\omega_d - \omega_0) - \gamma \sin \omega_d T\right]^2 + \gamma^2 (1 + \cos \omega_d T)^2}.
\label{R}
\ee
The transmittance in the long-time limit is given by $\mathcal{T} = \abssq{\alpha^{out}_B(\infty)}/\abssq{\alpha^{in}_A(t)} = 1 - \mathcal{R}$. From \eqref{gamma}, we see that the effective relaxation rate $\gamma$ also depends on the driving frequency $\omega_d$ (In the driven case, $\omega_0$ appearing in \eqref{gamma} needs to be replaced by $\omega_d$.). However, as long as the change of $\omega_d$ is small enough, \textit{i.e.}, $\Delta \omega_d \ll 2\pi/(N\tau) \sim \omega_0/N$, we can view $\gamma$ as approximately constant in the full range of driving frequencies. From \eqref{R}, we see that the condition for total reflection, $\mathcal{R} = 1$, is given by
\be
\omega_d = \omega_0 + \gamma \sin (\omega_d T),
\label{Ris1}
\ee
and the condition for total transmission, $\mathcal{R} = 0$, is given by $\omega_d T = (2n +1) \pi$, $n \in \Z$. In case the phase difference between the two legs is a multiple of $2\pi$, \textit{i.e.}, $\omega_0 T = 2n \pi$, \eqref{Ris1} has only one solution ($\omega_d = \omega_0$) for $\gamma T < 1$, but two additional solutions exist for $\gamma T > 1$.

\begin{figure}
\centering
\includegraphics[width=\linewidth]{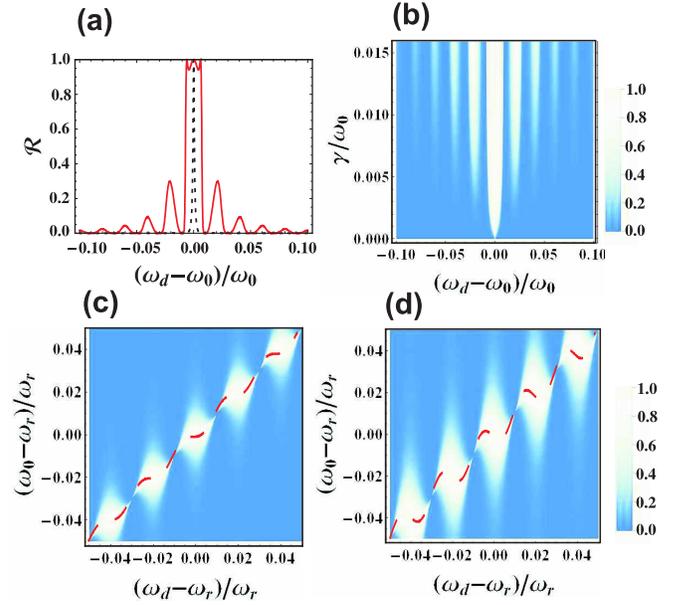}
\caption{\textbf{Reflectances}.
(a) Reflectance $\mathcal{R}$ as  a function of the scaled driving detuning $(\omega_d - \omega_0)/\omega_0$ for $\gamma T = 0.1$ (dashed black line) and $\gamma T = 2$ (red line).
(b) Reflectance $\mathcal{R}$ as  a function of the scaled detuning $(\omega_d - \omega_0)/\omega_0$ and the scaled decay rate $\gamma/\omega_0$.
(c) Reflectance $\mathcal{R}$ as a function of the scaled driving frequency detuning $(\omega_d - \omega_r)/\omega_r$ and the scaled transition frequency detuning $(\omega_0 - \omega_r)/\omega_r$ for $\gamma T = 1$, where $\omega_r$ is the reference frequency.
(d) The same as (c), but with $\gamma T = 2$. Other parameters: $\omega_0 T = 100 \pi$ for (a) and (b), $\omega_r T = 100 \pi$ for (c) and (d). \label{figR}}
\end{figure}

In \figref{figR}(a), we plot the reflectance $\mathcal{R}$ as a function of the scaled detuning $(\omega_d - \omega_0)/\omega_0$ for $\omega_0 T= 100 \pi$. The black curve corresponds to $\gamma T = 0.1$ and exhibits the typical features of a small atom: a single peak with perfect reflection at $\omega_d = \omega_0$. The red curve corresponds to $\gamma T = 2$ and shows some new features characteristic of the giant atom: (1) $\mathcal{R}$ exhibits a multi-peak structure as a function of detuning; (2) there are two additional driving frequencies resulting in $\mathcal{R} = 1$ beside the central peak at $\omega_d = \omega_0$. In fact, the multiple peaks correspond to the frequency peaks shown in \figref{figSpectrum}(c2). When the driving frequency $\omega_d$ resonates with one of the modes $\omega_{(k)}$, the reflectance shows a local maximum in \figref{figR}(a). However, the central peak at $\omega_d = \omega_0$ does not correspond to any mode $\omega_{(k)}$. This peak is  instead the result of a new pole in the complex plane due to the driving. The side peaks of $\mathcal{R} = 1$ can also be understood using \eqref{Ris1}. The transition frequency of the giant atom is shifted by $\gamma \sin(\omega_d T)$. Therefore, the driving frequency can be resonant with the shifted transition frequency of the giant atom again (if $\gamma T > 1$) when it deviates from $\omega_d = \omega_0$. In \figref{figR}(b), we plot the reflectance $\mathcal{R}$ as functions of the scaled detuning $(\omega_d - \omega_0)/\omega_0$ and the scaled decay rate $\gamma/\omega_0$ for $\omega_0 T= 100 \pi$.

In experiments, the transition frequency $\omega_0$ of the transmon qubit can be easily tuned \textit{in situ}; tuning the decay rate $\gamma$ is considerably more difficult. (As discussed below Eq.(\ref{gamma}), we have neglected the frequency dependence of the effective decay rate $\gamma$. Thus, the only way to tune the effective decay rate $\gamma$ is to change the single-finger decay rate $\gamma_0$ which is, however, fixed by the design of the device.) Thus, we plot the reflectance $\mathcal{R}$ as a function of both drive detuning $(\omega_d - \omega_r)/\omega_r$ and transmon detuning $(\omega_0 - \omega_r)/\omega_r$ in Figs.~{\ref{figR}}(c) and (d) for $\gamma T = 1$ and $\gamma T = 2$, respectively. Here, the reference frequency $\omega_r$ is given by $\omega_r = 100 \pi/T$. The dashed red curves correspond to the condition of total reflection given by \eqref{Ris1}.

\section{Two-phonon processes}
\label{sec:TwoPhononProcesses}

In this section, we study the scattering of a weakly coherent pulse from the giant atom, focusing on two-phonon processes. We construct the exact two-phonon scattering matrix following the diagrammatic approach of Ref.~\cite{Laakso2014} and use it to compute the leading-order contribution (in the phononic flux) to the second-order coherence functions and to the inelastic power spectrum of the scattered phonons. In addition, we find the first-order correction to the transmittance, extending the result of \secref{sec:SinglePhononScattering} beyond the single-phonon approximation.

\subsection{Two-phonon scattering matrix}

We assume that the phonon field is initially prepared in a coherent state in the form of a wavepacket centered around the driving frequency $\omega_d$. By defining a unitary operator
\be
U \equiv \exp \left[- i \omega_d t \left(\sp \sm + \sum_{\alpha = 1, 2} \int a^\dag_{\alpha \omega_p} a_{\alpha \omega_p} d\omega_p\right) \right],
\ee
the Hamiltonian in \eqref{Hstart} is transformed into  $H_{RF} \equiv U^\dag \mathscr{H} U + i \hbar \frac{dU^\dag}{dt} U$ in the frame rotating with frequency $\omega_d$. Dropping the fast oscillating terms in $H_{RF}$ and setting $\hbar=1$, we arrive at the Hamiltonian under the RWA,
\bea
H &=& - \frac{1 + \sigma_z}{2} \delta + \sum_{\alpha = 1, 2} \int d\omega \omega a^\dag_{\alpha \omega}a_{\alpha \omega} \nn\\
&+& \sum_{\alpha = 1, 2} \int d\omega \left(\nu_{\alpha \omega} a^\dag_{\alpha \omega} + \mathrm{H.c.}\right).
\label{Hv}
\eea
Here, we have defined the atomic detuning $\delta \equiv \omega_d - \omega_0$ and $\sz \equiv \ketbra{e}{e} - \ketbra{g}{g}$. The phonon frequency is also shifted by the driving frequency, \textit{i.e.}, $\omega \equiv \omega_p - \omega_d$. We have introduced the interacting operator $\nu_{\alpha \omega}$, which we call bare vertex operator,
\be
\nu_{\alpha \omega} \equiv \sqrt{\frac{\gamma}{4\pi}} \sum_{\beta = 1, 2} \sm e^{-i c_\alpha c_\beta (\omega T + \varphi)/2}.
\ee
The parameter $\varphi = \omega_d T = \omega_d L/v_g$ is the phase accumulated by the phonons during propagation from one leg to the other (we have used the relationship $k L = \omega_p T$ for SAWs).

We consider a rectangular pulse of spatial length $d$, initially created at a large distance ($\gg d$) to the left of the giant atom, propagating rightwards with a constant group velocity $v_g$. Keeping contributions from up to two phonons, we can write the initial phonon state
\bea
\ket{\Psi_i} &=& e^{-\bar{n}/2} e^{\sqrt{\bar{n}} b_1^\dag} \ket{0} \nn \\
&\approx& e^{-\bar{n}/2} \left[ \ket{0} + \sqrt{\bar{n}} b_1^\dag \ket{0} + \frac{\bar{n}}{2} b_1^{\dag \, 2} \ket{0} + O \left(\bar{n}^{3/2}\right) \right], \quad\quad
\label{psi0}
\eea
where $\bar{n} \ll 1$ is the mean number of phonons in the coherent state, and $b_1^\dag$ is a normalized wavepacket operator
\be
b_1^\dag = \int d \omega \phi (\omega) a_{1 \omega}^\dag, \quad
\phi (\omega) = \sqrt{\frac{2 v_g}{\pi d}} \frac{\sin(\omega d /2 v_g)}{\omega},
\ee
defined in terms of the plane-wave operator $a_{1 \omega}^\dag$ which creates a right-propagating phonon of frequency $\omega + \omega_d$ (the frequency $\omega$ in $a_{1 \omega}^\dag$ is measured from the driving frequency $\omega_d$). We also assume that the wavepacket $\phi (\omega)$ has a narrow bandwidth $\frac{2 \pi v_g}{d} \ll \gamma$, which implies that we can make the replacement $\phi (\omega) \to \sqrt{\frac{2 \pi v_g}{d}} \delta(\omega)$ whenever $\phi (\omega)$ is convolved with a function varying slowly on the bandwidth scale.

After scattering from the giant atom, the initial phonon state in \eqref{psi0} becomes the final one
\be
\ket{\Psi_f} = e^{-\bar{n}/2} \left[ \ket{0} + \sqrt{\bar{n}} S^{(1)} b_1^\dag \ket{0} + \frac{\bar{n}}{2} S^{(2)} b_1^{\dag \, 2} \ket{0} + O
\left(\bar{n}^{3/2}\right) \right],
\label{psi}
\ee
where $S^{(1)}$ and $S^{(2)}$ are the one- and two-phonon scattering operators. They can be expressed as \cite{Laakso2014}
\bea
&& S^{(1)} = \left\{ \delta_{s' s} - 2 \pi i P_- v_{s'} M (\omega) v^\dag_{s} P_- \delta_{\omega' \omega} \right\} a^\dag_{s'} a_{s}, \label{S1} \\
&& S^{(2)} = \left\{ \frac12 \delta_{s'_1 s_1} \delta_{s'_2 s_2} - 2 \pi i P_- v_{s'_1} M (\omega'_1) \left[ \delta_{\omega'_1 \omega_1} \delta_{s'_2 s_2} + \right. \right. \nonumber \\
&& \left. \left. + W_{s'_2 s_2} (E) M (\omega_1) \delta_{\omega'_1 + \omega'_2, \omega_1 + \omega_2} \right] v^\dag_{s_1} P_- \right\} a_{s'_1}^\dag a_{s'_2}^\dag a_{s_2} a_{s_1}, \label{S2}
\eea
where $s = (\alpha, \omega)$ is a multi-index, and we implicitly assume summation/integration over it, when it is repeated. The parameter $E$ must eventually be set to the value of an incoming-state energy $\omega_1 + \omega_2$; for the initial state (\eqref{psi0}) in our convention about the energy reference point it equals $E = 0$. The Green's functions of the qubit in the ground $G(E) = \frac{P_-}{E + i \eta}$ and excited $M (E) = \frac{P_+}{E + \delta + \Sigma(E)}$ states are spanned by the corresponding projectors $P_{\pm} = \sigma_{\pm} \sigma_{\mp} = \frac{1 \pm \sz}{2}$. The self-energy of the ground state is infinitesimally small ($\eta \to 0^+$), while the self-energy of the excited state $\Sigma(E)$ has to be established. The effective two-phonon vertex $W_{s'_2 s_2} (E)$ also requires specification.

We note that the Hilbert space of the qubit is two-dimensional, and, in addition, the Hamiltonian in \eqref{Hv} is written in the RWA. This means that in a diagrammatic representation of $\Sigma(E)$ and $W_{s' s}(E)$, the bare vertices $v$ and $v^\dag$ must alternate each other. Along with an application of Wick's theorem, this leads to the following exact equations:
\bea
\Sigma (E) &=& v_s^\dag G (E - \omega) v_s , \label{self} \\
W_{s's} (E) &=& w_{s's} (E) + w_{s' s_1} (E) M (E - \omega_1) W_{s_1 s} (E), \label{Wss}
\eea
where $w_{s's} (E) = v_s^\dag G (E - \omega - \omega') v_{s'}$, and \eqref{Wss} is obtained from the iteration $w + wMw +
\ldots$.

A simple calculation shows that \eqref{self} yields $\Sigma (E) = - i P_+ \gamma \left[1 + e^{i (E T + \varphi)} \right]$, and this is sufficient to recover the single-phonon scattering matrix. Thus, we obtain
\be
S^{(1)} b_1^{\dagger} \ket{0} =  s_{\alpha'_1, 1} \phi (\omega'_1) a_{\alpha'_1, \omega'_1}^{\dagger} \ket{0},
\ee
where the matrix elements
\bea
s_{1,1} &=& \frac{\delta - \gamma \sin \varphi}{\delta + i \gamma (1+ e^{i \varphi})},\label{s1,1} \\
s_{2,1} &=& -i \gamma \frac{1 + \cos \varphi}{\delta + i \gamma (1 + e^{i \varphi})}.\label{s2,1}
\eea
Note that the scattering matrix elements $s_{1,1}$ and $s_{2,1}$ connect the incident right-propagating phonons to the scattered right-propagating and left-propagating phonons, respectively. Therefore, the reflectance and transmittance are given by $\mathcal{T}=|s_{1,1}|^2$ and $\mathcal{R}=|s_{2,1}|^2$, which coincides with the definitions found in the previous section and in \appref{app:SinglePhononRTCoefficients}.

The effective two-phonon vertex $W_{s's} (E)$ obeying \eqref{Wss} accounts for multiple excursions of two correlated phonons between the two legs. Parameterizing
\be
W_{s's} (E) = \sum_{\beta', \beta} \frac{\gamma P_+}{4 \pi} e^{-i c_{\alpha'} c_{\beta'} (\omega' T + \varphi)/2} e^{i c_{\alpha} c_{\beta} (\omega T + \varphi)/2} \overline{W} (\omega',\omega),
\ee
we simplify \eqref{Wss} down to
\bea
\overline{W} (\omega', \omega) &=& \frac{1}{E - \omega' - \omega + i \eta} \nn \\
&+& \frac{\gamma}{2\pi} \int d\omega_1 \frac{1}{E - \omega' - \omega_1 + i \eta} \nn \\
&&\times \frac{e^{i \omega_1 T + i \varphi}}{E - \omega_1 + \delta + i \gamma} \overline{W} (\omega_1, \omega).
\label{intW}
\eea

The first term on the RHS of \eqref{intW} does not depend on the travel time $T = L/v_g$ between the legs. The function $\overline{W} (\omega', \omega)$ is analytic in the lower half-plane in both its arguments. In the regime $\gamma T \ll 1$, we can neglect the term $e^{i \omega_1 T}$ under the integral and close the integration contour in the lower half-plane, which leads to the second term vanishing. Therefore, at short inter-leg distances we can approximate $\overline{W} (\omega', \omega) \approx \frac{1}{E - \omega' - \omega + i \eta}$, identifying this term with the Markovian contribution.

For larger separations $\gamma T \gtrsim 1$, we need the full solution of \eqref{intW}. Taking into account the particular form of the initial state in \eqref{psi0}, it suffices to solve \eqref{intW} for $E = \omega = 0$. This can be done analytically, and we find
\bea
\overline{W} (-q, 0)|_{E=0} &=& \frac{1}{q + i \eta} + F (q), \\
F (q) &=& -\frac{ i \gamma e^{i \varphi} }{\lambda + i \gamma e^{i \varphi}} \sum_{\sigma = \pm, 0} C_{\sigma} \frac{e^{i q T} - e^{-i \sigma p T}}{q + \sigma p},
\eea
where
\bea
p &=& \sqrt{\lambda^2 + \gamma^2 e^{2 i\varphi}}, \quad
\lambda =\delta+i\gamma, \\
C_{\pm} &=& \pm \frac{(\pm p - \lambda) e^{\pm i p T} - i \gamma e^{i \varphi}}{2 (p \cos p T - i \lambda \sin p T)}, \quad
C_0 = -1.
\label{C}
\eea
The parameter $p$ appears only in the non-Markovian part $F (q)$ of the term (32). Its real and imaginary parts contain a new oscillation frequency $\textup{Re} (p)$ and a new relaxation rate $\textup{Im} (p)$, which can only manifest themselves in the two-photon inelastic scattering processes in the non-Markovian regime $|p| \cdot T \gg 1$.

Finally, we derive
\bea
\frac12 S^{(2)} b_1^{\dag \, 2} \ket{0} &=& \sum_{\alpha'_1, \alpha'_2} \frac12 s_{\alpha'_1, 1} s_{\alpha'_2, 1} \phi (\omega'_1) \phi (\omega'_2) a_{\alpha'_1, \omega'_1}^\dag a_{\alpha'_2, \omega'_2}^\dag \ket{0} \nn \\
&+& \sum_{\alpha'_1, \alpha'_2} \frac{2 v_g \gamma}{d} s_{2,1} \int dq \cos \frac{q T + \varphi}{2} \cos \frac{-q T + \varphi}{2} \nn \\
&&\times M (q) \left[ \frac{1}{q} + F (q) \right] a_{\alpha'_1, q}^\dag a_{\alpha'_2, -q}^\dag \ket{0},
\label{s2act2}
\eea
thus completing the determination of the scattering state in \eqref{psi}. In this expression (and in the following), we use the notation $M (q)$ for $\frac{1}{q + \lambda + i \gamma e^{i q T + i \varphi}}$, omitting the associated matrix structure.

In the Markovian regime $|p| \cdot T \ll 1$, the second -- inelastic -- contribution to Eq.~(\ref{s2act2}) is approximated by
\begin{align*}
\sum_{\alpha'_1 , \alpha'_2} \frac{2 v_g \gamma}{d} s_{2,1} \cos^2 \frac{\varphi}{2} \int dq \frac{a_{\alpha'_1, q}^{\dagger}  a_{\alpha'_2, -q}^{\dagger} }{(q+ \lambda + i \gamma e^{i \varphi}) q} | 0 \rangle,
\end{align*}
where the $q$-dependence of the integrand features only simple poles. We also note the absence of the parameter $p$ in this expression.

\subsection{Correction to the transmittance}
\begin{figure}
\centering
\includegraphics[width=\linewidth]{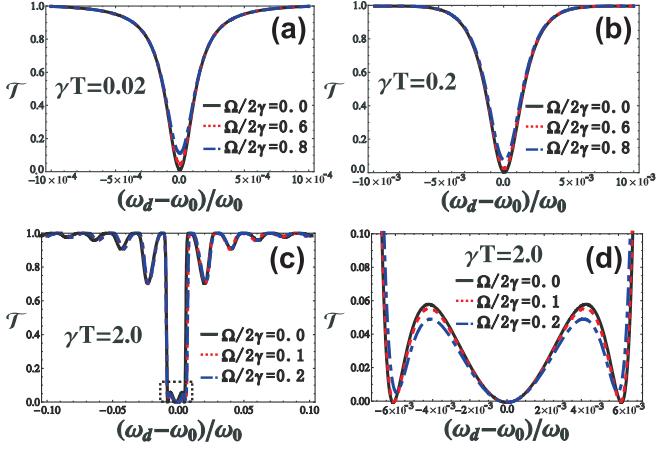}
\caption{\textbf{Transmittance including two-phonon processes.}  The plots show $\mathcal{T} = \abssq{s_{1,1}+\delta s_{1,1}}$ as a function of normalized driving detuning $(\omega_d - \omega_0)/\omega_0$. In all figures, $\omega_0 T = 100 \pi$.
(a) $\mathcal{T}$ for $\gamma T = 0.02$ and various scaled driving parameters $\Omega/(2\gamma)$.
(b) Same as (a), but with $\gamma T = 0.2$.
(c) $\mathcal{T}$ for $\gamma T = 2.0$ and various scaled driving parameters $\Omega/(2\gamma)$.
(d) Zoom-in of the dashed box in (c).
\label{fig:transmittance}}
\end{figure}

Knowing the exact two-photon S-matrix (\eqref{S2}), we can find the first nonlinear correction to the transmittance. We calculate $\brakket{\Psi_f}{a_1 (x)}{\Psi_f}=e^{i \omega_0 x/v_g} \sqrt{f} [s_{1,1} + \delta s_{1,1} + O (f^2)]$, where $f = \bar{n} v_g/d$ is a phonon flux. To compare our results for a giant atom to known results for a small atom, we introduce the driving amplitude $\Omega\equiv \sqrt{8\gamma f}$ (Rabi frequency in the small-atom limit $\gamma T \rightarrow 0$), which is widely used in the study of quantum optics. Then, we write the correction $\delta s_{1,1}$ to the single-phonon transmittance
\be
\delta s_{1,1} = \frac{1}{2}\left(\frac{\Omega}{2\gamma}\right)^2 \frac{8 i p \gamma^3 \cos^4 \frac{\varphi}{2} (\lambda \cos p T - i p \sin p T + i \gamma e^{i \varphi})}{\abssq{\lambda + i \gamma e^{i \varphi}} (\lambda + i \gamma e^{i \varphi})^2 (p \cos p T - i \lambda \sin p T)} .
\label{deltat}
\ee
While the linear transmittance $s_{1,1}$ corresponds to the transition $\ket{1} \to \ket{0}$ (elimination of a single phonon), the correction $\delta s_{1,1}$ corresponds to a measurement of a phonon in a two-phonon state, $\ket{2} \to \ket{1}$. Let us analyze $\delta s_{1,1}$ in different limiting cases.

For $\gamma T \ll 1$, we obtain
\be
\delta s_{1,1} \approx  \frac{1}{2}\left(\frac{\Omega}{2\gamma}\right)^2 \frac{8 i \gamma^3 \cos^4 \frac{\varphi}{2} }{\abssq{\lambda + i \gamma e^{i \varphi}} (\lambda + i \gamma e^{i \varphi})} .
\label{dt_smR}
\ee
In the small-atom limit $\varphi\rightarrow 0$, we can furthermore simplify the above correction
\be
\delta s_{1,1} \approx  \frac{1}{2}\left(\frac{\Omega}{2\gamma}\right)^2 \frac{1+i\frac{\delta}{2\gamma} }{\left[1+\left(\frac{\delta}{2\gamma}\right)^2\right]^2},
\label{}
\ee
which is consistent with the result in Ref.~\cite{Astafiev2010} for the study of a small artificial atom (recall that the total relaxation rate of our atom in this limit is $2\gamma$). 

For $\gamma T \gg 1$ and $p \neq 0$ ($\mathrm{Im} \, p > 0$), we find that
\be
\delta s_{1,1} \approx  \frac{1}{2}\left(\frac{\Omega}{2\gamma}\right)^2 \frac{8 i \gamma^3 \cos^4 \frac{\varphi}{2}}{\abssq{\lambda + i \gamma e^{i \varphi}} (\lambda + i \gamma e^{i \varphi})} \frac{p}{\lambda + i \gamma e^{i \varphi}},
\label{dt_laR}
\ee
which differs from its short-distance counterpart in \eqref{dt_smR} by the additional factor at the end.

For $p = 0$ (resonance $ \delta=0$ and bright state $\varphi=2k\pi, k\in \mathbb{N}$), we derive the expression directly from \eqref{deltat}
\be
\delta s_{1,1} =  \frac{1}{2}\left(\frac{\Omega}{2\gamma}\right)^2 \frac{1}{1 + \gamma T} ,
\label{dt_spec}
\ee
valid for arbitrary $T$. Since in this case $s_{1,1}= 0$, \eqref{dt_spec} represents the leading contribution to the transmittance. Unlike \eqref{dt_laR}, it vanishes at large $T$.

We plot the corrected transmittance $\mathcal{T} = \abssq{s_{1,1} + \delta s_{1,1}}$ in \figref{fig:transmittance} for different parameter regimes. Fig.~\ref{fig:transmittance}(a) and (b) show the transmittance for $\gamma T= 0.02$ and $\gamma T = 0.2$, respectively. We see that the two-phonon process basically enhances the transmittance. The reason is that the atom can only interact with a single phonon at any given time. When $\gamma T$ is small, corresponding to the limit of a small atom with Markovian dynamics, a second incoming phonon will thus not interact with the atom and simply be transmitted forward \cite{Shen2007, Chang2007, Hoi2012}. However, for larger $\gamma T=2.0$, when the Markov approximation breaks down, the transmittance through the giant atom shows a more complicated structure as can be seen in Figs.~\ref{fig:transmittance}(c) and (d). In particular, we observe that the two-phonon process does not always enhance the transmttance, but instead sometimes suppresses it. This is due to the fact that the giant atom can interact with one phonon while the other phonon travels between the two connection points. These two phonons can then interfere constructively or destructively in a way that is not possible with a small atom.

\subsection{Inelastic power spectrum}

\begin{figure}
\centering
\includegraphics[width=\linewidth]{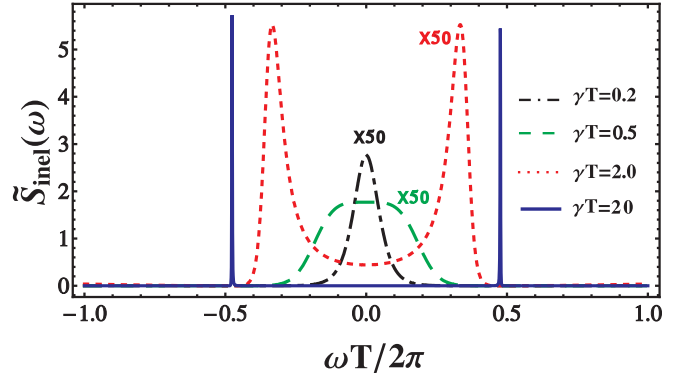}
\caption{{\bf Inelastic power spectra.} The black dotted dashed, green long dashed, red short dashed, and blue solid curves are the scaled inelastic power spectra $\tilde{S}_{\mathrm{inel}}(\omega)\equiv({2\gamma}/{\Omega})^4 S_{\mathrm{inel}}(\omega)$ for $\gamma T=0.2$ (enlarged $50$ times), $\gamma T=0.5$ (enlarged $50$ times), $\gamma T=2.0$ (enlarged $50$ times) and $\gamma T=20$, respectively. Other parameters: $\delta/\omega_0=0.0$, $\varphi=2k\pi, k\in Z^+$. \label{SinelgammaT}}
\end{figure}

At weak coherence $\bar{n} \ll 1$, the elastic scattering dominates over the inelastic scattering: the former receives the leading $O (\Omega^2)$ contribution from a single-phonon process, while the latter starts to happen when at least two phonons are involved. This gives the $O (\Omega^4)$ contribution.

Let us compute these leading terms in the power spectrum of the giant atom in the state from \eqref{psi}. We consider $g_{\alpha}^{(1)} (\tau) = \brakket{\Psi_f}{a_{\alpha}^\dag (x - v_g \tau) a_{\alpha} (x)}{\Psi_f}$, where $a_{\alpha} (x) = \sqrt{\frac{v_g}{2 \pi}} \int d\omega a_{\alpha \omega} e^{i(\omega + \omega_0) x/v_g}$, and establish
\bea
g_{\alpha}^{(1)} (\tau) &=& e^{i \omega_0 \tau} \left\{ (8\gamma)^{-1}\Omega^2 \abssq{s_{\alpha , 1}} -(4\gamma)^{-2}\Omega^4 \mathcal{R} \mathrm{Im} \left[s_{\alpha ,1}^*     \Lambda M (0) \right] \right\} \nn \\
&+& \int d\omega e^{i (\omega_0 + \omega) \tau} S_{\mathrm{inel}} (\omega) + O \left(\Omega^6\right),
\label{g1mikhail}
\eea
where $\mathcal{R} = \abssq{s_{2,1}}$ and
\be
\Lambda = 1 + \frac{i \gamma e^{i \varphi}}{p} \left[C_+ \left(1 - e^{-i p T} \right) - C_ - \left(1 - e^{i p T} \right) \right] .
\label{Lambda}
\ee
The leading inelastic contribution is given in terms of the corresponding power spectrum
\bea
&&S_{\mathrm{inel}} (\omega) = \frac{ \Omega^4}{4\pi}\mathcal{R} \cos^2 \frac{\omega T + \varphi}{2} \cos^2 \frac{-\omega T + \varphi}{2} \nn \\
&&\times \abssq{\frac{M (\omega) - M ( - \omega)}{2 \omega} + \frac{M (\omega) F (\omega) + M ( -\omega) F (-\omega)}{2}}. \quad
\label{Sinel}
\eea
We see that the inelastic power is on the order of $O (\Omega^4)$.

It is important to check the power conservation. Summing $g_{\alpha}^{(1)} (0)$ over the channels $\alpha$ and using the unitarity of the single-phonon matrix from \eqref{S1}, we obtain the incoming power up to leading order $O (\Omega^2)$. Therefore, the elastic and inelastic contributions to the power in $O (\Omega^4)$ must cancel each other, \textit{i.e.,}
\be
- (4\gamma)^{-2}\Omega^4 \mathcal{R} \sum_{\alpha} \mathrm{Im} \left[s_{\alpha, 1}^* \Lambda M (0) \right] + 2 \int d\omega S_{\mathrm{inel}} (\omega) = 0.
\label{powerconsevation}
\ee
This relationship indeed holds due to the unitarity of the two-phonon scattering matrix in \eqref{S2}.  In the resonant case $\delta/\omega_0=0.0$ and for the bright state $\varphi=2k\pi, k\in Z^+$, we have $M(0)=-i(2\gamma)^{-1}$ and $\Lambda=1/(1+\gamma T)$ from Eqs.~(\ref{C}) and (\ref{Lambda}). From Eqs.~(\ref{s1,1}) and (\ref{s2,1}), we have the scattering matrix elements $s_{2,1}=-1$ and $s_{1,1}=0$. The total power of the inelastic spectrum can be calculated from \eqref{powerconsevation}:
\be
 \int d\omega S_{\mathrm{inel}}(\omega) = \Big(\frac{\Omega}{2\gamma}\Big)^4\frac{\gamma}{4(1+\gamma T)}.
\label{totalinelpower}
\ee
 In the small-atom limit $\gamma T\rightarrow 0$, the total power of the inelastic spectrum is $\frac{\gamma}{4}\Big(\frac{\Omega}{2\gamma}\Big)^4$. However, in the giant-atom limit $\gamma T\gg 1$, the total inelastic power is $\frac{1}{4T}\Big(\frac{\Omega}{2\gamma}\Big)^4$. In Fig.~\ref{SinelgammaT}, we plot the scaled inelastic power spectrum $\tilde{S}_{\mathrm{inel}}(\omega)\equiv({2\gamma}/{\Omega})^4 S_{\mathrm{inel}}(\omega)$ as a function of the dimensionless frequency $\omega T/2\pi$. Due to the narrow bandwidths and the high peaks in the large-atom limit $\gamma T=20$, we enlarge the plots for $\gamma T=0.2$, $\gamma T=0.5$ and $\gamma T=2.0$ by fifty times.

For point-like atoms ($T=0$), we can simplify the inelastic power spectrum \eqref{Sinel} on resonance $\delta/\omega_0=0$ and bright state $\varphi=2k\pi, k\in Z^+$
\bea
S_{\mathrm{inel}} (\omega) = \frac{1}{4\pi}\left(\frac{\Omega}{2\gamma}\right)^4 \left(\frac{4\gamma^2}{\omega^2+4\gamma^2}\right)^2.
\label{SinelSmall}
\eea
This gives a single peak around the central resonant frequency in the elastic spectrum as shown by the black dot-dashed line in Fig.~\ref{SinelgammaT}. However, if we increase the size of the atom, the central peak will split into two peaks.
In the general case $\gamma T>0$, we have the inelastic power spectrum
\bea
&&S_{\mathrm{inel}} (\omega) = \nonumber\\
 &&\frac{ \Omega^4 }{16\pi(1+\gamma T)^2}
\left[\frac{1+\cos\omega T}{(\omega-\gamma\sin\omega T)^2+\gamma^2(1+\cos\omega T)^2}\right]^2. \nonumber\\
\label{SinelGiant}
\eea
The critical point where the central peak splits into two peaks can be determined by the second derivative of the inelastic power spectrum, i.e., $\frac{d^2}{d\omega^2}S_{\mathrm{inel}} (\omega)|_{\omega=0}=0$, which gives the critical value $(\gamma T)_c=1/2$. The mechanism of two peaks appearing here is different from that of two side peaks in the famous Mollow triplet \cite{Mollow1969}, which comes from relatively large driving $\Omega/(2\gamma)>1$ \cite{Mollow1969,Milburn2008}. In fact, our two-phonon expansion is only valid for $\Omega/(2\gamma) < 1$. The two peaks found here come from the time delay of the giant atom, not from strong driving.

\subsection{Production of phonon pairs}

To further understand the meaning of the inelastic power spectrum $S_{inel}(\omega)$, we rewrite the scattered two-phonon state from \eqref{s2act2} as
\bea
 S^{(2)} b_1^{\dag \, 2} \ket{0} &=& \frac{2\pi v_g }{d}\sum_{\alpha_1, \alpha_2}  s_{\alpha_1, 1} s_{\alpha_2, 1}  a_{\alpha_1, 0}^\dag a_{\alpha_2, 0}^\dag \ket{0} \nn \\
&+&\int d\omega\psi(\omega)\Big( a_{1, \omega}^\dag a_{1, -\omega}^\dag + a_{2, \omega}^\dag a_{2, -\omega}^\dag\Big)\ket{0} \nn \\
&+&\int d\omega\psi(\omega)\Big( a_{1, \omega}^\dag a_{2, -\omega}^\dag + a_{2, \omega}^\dag a_{1, -\omega}^\dag\Big)\ket{0}.\ \ \
\label{s2final}
\eea
Here we have used the narrow bandwidth assumption, \textit{i.e.}, $\phi(\omega)=\sqrt{\frac{2\pi v_g }{d}}\delta(\omega)$.
The first line on the RHS of \eqref{s2final} means that the two phonons travel through the transmission line independently. The second and third lines on the RHS of \eqref{s2final} represent the final states of two phonons after inelastic scattering (exchanging energy). Due to energy conservation, the two scattered phonons are always generated in pairs with frequencies of opposite signs (with reference to the driving frequency $\omega_d$). The second line on the RHS of \eqref{s2final} represents the superposition state of a right-propagating phonon pair and a left-propagating phonon pair.  The third line on the RHS of \eqref{s2final} represents the phonon pair of a right-propagating phonon and a left-propagating phonon. The coefficient of these photon-pair states at frequency $\omega$ is given by
\bea
&&\psi(\omega)\equiv s_{2,1}\frac{2 v_g \gamma}{d} \cos \frac{\omega T + \varphi}{2} \cos \frac{-\omega T + \varphi}{2}\nn \\
&& \times\Big[\frac{M (\omega) - M ( - \omega)}{2 \omega} + \frac{M (\omega) F (\omega) + M ( -\omega) F (-\omega)}{2}\Big].
\label{psiomega}
\eea
Compared to \eqref{Sinel}, we see that the inelastic power spectrum $S_{inel}\propto|\psi(\omega)|^2$ is a direct measure of the production of phonon pairs with frequency $\omega$.

In \figref{SinelgammaT}, we plot the scaled inelastic power spectrum $\tilde{S}_{inel}(\omega)$ with the parameters $\delta/\omega_0=0$ and $\varphi=2k\pi$, $k\in Z^+$, which corresponds to perfect reflection $\mathcal{R}=|s_{2,1}|^2=1$ and zero transmission $\mathcal{T}=|s_{1,1}|^2=0$ in the single-phonon approximation. In this case, no phonons are transmitted by elastic scattering, \textit{i.e.}, the first line on the RHS of \eqref{s2final} gives no contribution to the transmission channel. However, phonons are still allowed to transmit through inelastic scattering as described by the second line on the RHS of \eqref{s2final} (two phonons are transmitted) and the third line on the RHS of \eqref{s2final} (one phonon is transmitted and the other is reflected). From \figref{SinelgammaT}, we see that, in the small-atom limit $\gamma T=0.2$, the production of phonon pairs centres at $\omega=0$ decreasing with frequency on both sides. However, when the atom becomes larger, \textit{e.g.}, $\gamma T=2$, the phonon pair production centres around two well-separated frequency regions. For the giant atom with $\gamma T=20$, we generate phonon pairs of frequencies $\omega\approx \pm \pi/T$ with a narrow bandwidth.

\begin{figure}
\centering
\includegraphics[width=\linewidth]{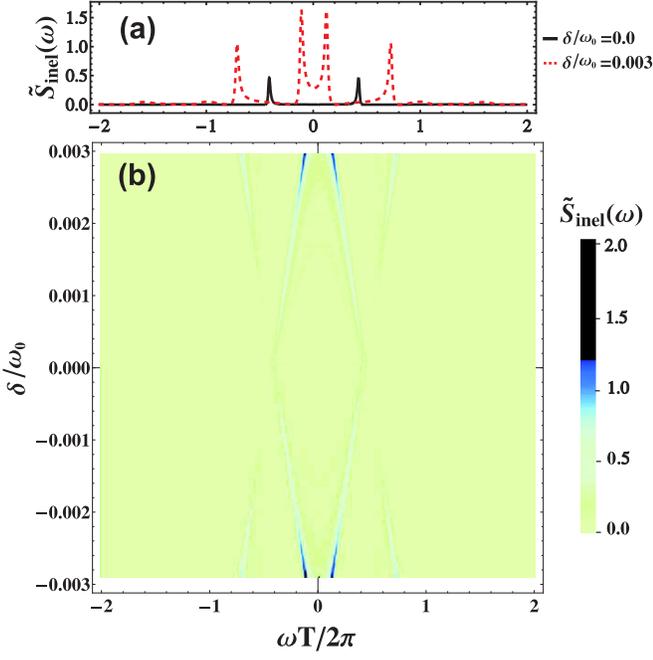}
\caption{{\bf Inelastic power spectra vs detunings.} (a) The black solid and red dashed curves are the scaled inelastic power spectra $\tilde{S}_{inel}(\omega)$ for $\delta/\omega_0=0$ and $\delta/\omega_0=0.003$, respectively. (b) Density plot of $\tilde{S}_{inel}(\omega)$ with the detuning $\delta/\omega_0 \in [-0.003,0.003]$. Other parameters: $\gamma T=5.0$, $\varphi=200\pi$ for both figures.
\label{Sineldetuning}}
\end{figure}

We further plot the scaled inelastic power spectrum as a function of the dimensionless driving detuning $\delta/\omega_0$ in \figref{Sineldetuning}. We set the parameters as $\gamma T=5.0$ and $\varphi=200\pi$ and change the detuning value continuously from $\delta/\omega_0=-0.003$ to $\delta/\omega_0=0.003$ in \figref{Sineldetuning}(b). It is shown that the inelastic power spectrum is symmetric with respect to the detuning. At zero detuning, $\delta/\omega_0=0$, we see two peaks in the inelastic power spectrum as shown by the black curve in \figref{Sineldetuning}(a). For a small detuning $\delta/\omega_0=0.003$, however, we see four peaks as shown by the red curve in \figref{Sineldetuning}(a). This means we can generate phonon pairs at two different frequencies. From the third line on the RHS of \eqref{s2final} and \eqref{psiomega}, we see that the production of phonon pairs is proportional to the reflectance $\mathcal{R}=|s_{2,1}|^2$. Note that at $\varphi = \pi + 2 \pi k, \ k\in\Z^+$, the backward scattering is absent ($\mathcal{R} = 0$), and the inelastic scattering vanishes -- all phonons propagate forward without any obstruction due to the formation of a dark state in the giant atom. In this case, as a result, no phonon pairs are generated.

\ \

\ \

\subsection{Second-order coherence correlation functions}

A further important quantity is
the second-order coherence correlation function defined by
\be
G_{\alpha' \alpha}^{(2)} (\tau) \equiv \brakket{\Psi_f}{a_{\alpha}^\dag (x) a_{\alpha'}^\dag (x - v_g \tau) a_{\alpha'} (x - v_g \tau) a_{\alpha} (x)}{\Psi_f }.
\label{G2Mikhal}
\ee
This correlation function corresponds to the probability of detecting a phonon in channel $\alpha$  at time $\tau$ after detecting a first one in channel $\alpha'$ for the final state $|\Psi_f\rangle$ (see, \textit{e.g.}, \cite{Milburn2008,Dorner2002}). It is also useful to introduce the normalized second-order coherence correlation function
\be
g_{\alpha' \alpha}^{(2)} (\tau) = \frac{\brakket{\Psi_f}{a_{\alpha}^\dag (x) a_{\alpha'}^\dag (x - v_g \tau) a_{\alpha'} (x - v_g \tau) a_{\alpha} (x)}{\Psi_f }}{g^{(1)}_{\alpha'} (0) g^{(1)}_{\alpha} (0)}.
\label{g2byMikhal}
\ee
In the two-phonon approximation, it becomes
\be
g_{\alpha' \alpha}^{(2)} (\tau) = \abssq{1 + \kappa_{\alpha' \alpha} (\varphi) \left[ \cos \varphi I_0 (\tau) + I_1 (\tau) \right]},
\label{g2aa}
\ee
where the coefficients
\bea
\kappa_{11} (\varphi) &=& \frac{ \gamma^2 (1+\cos \varphi)}{(\delta - \gamma \sin \varphi )^2}, \\
\kappa_{22} (\varphi) &=& - \frac{1}{1+\cos \varphi}, \\
\kappa_{12} (\varphi) &=& \kappa_{21} (\varphi) = \frac{i \gamma}{\delta - \gamma \sin \varphi}
\eea
specify the coherence correlations in the transmitted ($g_{11}^{(2)}$) and reflected ($g_{22}^{(2)}$) channels, as well as the cross-correlations ($g_{12}^{(2)} = g_{21}^{(2)}$).

The delay-time dependence enters in \eqref{g2aa} via the functions
\bea
I_0 (\tau) &=& \frac{M^{-1} (0)}{ \pi i} \int d q M (q) \left( \frac{1}{q} + F (q) \right) \cos q \tau, \\
I_1 (\tau) &=& \frac{I_0 (\tau - T) + I_0 (\tau + T)}{2} .
\eea
Performing the integral, we obtain them in the explicit form
\bea
&& I_0 (\tau) = \frac{1}{2 (p \cos p T - i \lambda \sin p T)} \nn \\
&\times& \{ e^{-i p (\abs{\tau} + T)} M^{-1} (p) - e^{i p (\abs{\tau} + T)} M^{-1} (-p) \nn \\
&+& \sum_{n=0}^{\infty} \Theta (\abs{\tau} - n T ) [g_n^{(-)} (\tau) - g_n^{(+)} (\tau)] \},
\label{i0fin}
\eea
where
\bea
g_n^{(\pm)} (\tau) &=& (- i \gamma e^{i \varphi})^n e^{\mp i p T} \frac{( \pm p + \lambda + i \gamma e^{i \varphi} e^{\pm i p T})^2}{(\pm p +\lambda)^{n+1}} \nn \\
&\times& \left[ e^{\mp i p (|\tau | - n T)} - e^{i \lambda (\abs{\tau} - n T) } f_n^{(\pm)} (\tau) \right] , \\
f_n^{(\pm)} (\tau) &=& \sum_{m = 0}^n \frac{[- i (\abs{\tau} - n T) (\pm p + \lambda )]^{m}}{m!}.
\eea

For small $T$, \textit{i.e., $\gamma T \ll 1$,} we can neglect the explicit $T$ dependence, keeping only $\varphi$ finite. Thus, we obtain
\be
I_0 (\tau) \approx I_1 (\tau) \approx e^{i (\lambda + i \gamma e^{i \varphi}) \tau},
\ee
which leads to
\be
g_{22}^{(2)} (\tau) = \abssq{ 1 - e^{i (\delta -\gamma \sin \varphi)\tau} e^{- \gamma ( 1 + \cos \varphi )\tau} } .
\label{g22small}
\ee
The small atom can only absorb and emit one phonon at a time, which results in $g_{22}^{(2)} (0)=0,$ exhibiting the typical antibunching behaviour of a single-phonon state.

For large $T$, \textit{i.e.}, $\gamma T \gg 1$, we neglect the terms containing $\abs{e^{i p T}} \ll 1$ (without loss of generality we assume $\mathrm{Im} \, p >0$, since \eqref{i0fin} is invariant under $p \to -p$). For $\tau > 0$, this yields
\bea
I_0 (\tau) &\approx& \sum_{n = 0}^{\infty} \Theta ( n T < \tau < (n+1) T) (- i \gamma e^{i \varphi})^n \nn \\
&\times& \left[ \frac{e^{i p (\tau - n T)}}{(-p + \lambda)^{n}} - i \gamma e^{i \varphi} \frac{e^{-i p (\tau - (n+1) T)}}{(p + \lambda)^{n+1}} \right. \nn \\
&& \left. - e^{i \lambda (\tau- n T)} \left( \frac{f_n^{(-)} (\tau)}{(-p + \lambda)^{n}} - \frac{f_n^{(+)} (\tau) }{(p + \lambda)^{n}}\right) \right],
\label{i0app}
\eea
\textit{i.e.}, in a given interval $[n T, (n+1) T]$ there are terms which exponentially decay inward the interval at rate $\mathrm{Im} \, p$ from both ends, and in addition there is a term exponentially decaying at rate $\gamma$ from the left end and multiplied by a polynomial of a degree $n-1$.

\begin{figure}
\centering
\includegraphics[width=\linewidth]{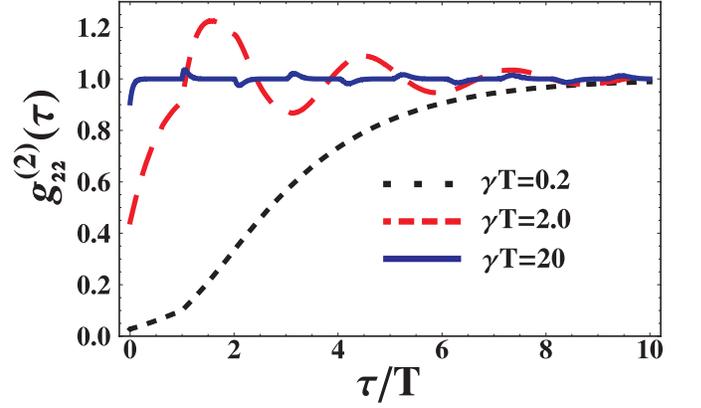}
\caption{{\bf Second-order correlation function for phonons reflected from a giant atom.} The plot shows the correlation function $g_{22}^{(2)} (\tau)$ for three choices of $\gamma T$, taken from regions $B$, $C$, and $D$ of the parameter space shown in \figref{fig:ParameterRegimes}. For the other parameters, we have chosen the resonance condition $\delta/\omega_0 = 0$ and the constructive-interference condition $\omega_0 T = 2 k \pi$, $k \in \Z$. \label{g2function}}
\end{figure}

There is, however, a special case $p \to 0$, in which the approximation of \eqref{i0app} is not applicable. It is realized when $\delta = 0$ and $\varphi = 2 \pi k, k \in \Z$. For these parameters a single phonon is fully reflected, so we focus on the correlation function for two phonons in the reflected channel as well. Its exact expression reads
\bea
g_{22}^{(2)} (\tau) &=& \left( 1 + \frac{1}{1 + \gamma T} \sum_{n = 0}^{\infty} \Theta (\tau - n T) K_n (\tau - n T) \right)^2, \\
K_n (z) &=& (-1)^{n+1} e^{-\gamma z} \frac{(\gamma z)^n}{n!} .
\label{Kn}
\eea
In \figref{g2function}, we plot the this correlation function for different choices of $\gamma T$. For $\gamma T \ll 1$ (short dashed black line), we see that the behaviour is close to that of a small atom, which displays perfect antibunching ($g_{22}^{(2)}(0) = 0$) on resonance \cite{Chang2007, Hoi2012}. For larger values of $\gamma T$ (long dashed red and solid blue lines), there is a possibility that a photon was emitted from the right leg of the giant atom at an earlier time, which results in $g_{22}^{(2)}(0) \neq 0$.

For $\gamma T \gg 1$, each contribution $K_n (\tau - n T)$ is localized in the beginning of the corresponding interval, being exponentially small at its end. Therefore, the contributions from different intervals do not overlap, and $g_{22}^{(2)}$ is represented by a sequence of kinks attached to the line of unity height. The shape of the $n$th kink ($n \geq 1$) is given by
\be
g_{22}^{(2)} (n T + z) - 1 \approx \frac{2}{\gamma T} K_n (z).
\ee
Because of the sign factor in \eqref{Kn}, every odd kink rises upwards, while every even kink dips downwards, and we observe an alternation of bunching and antibunching properties of the reflected phonons.

\begin{figure}
\centering
\includegraphics[width=\linewidth]{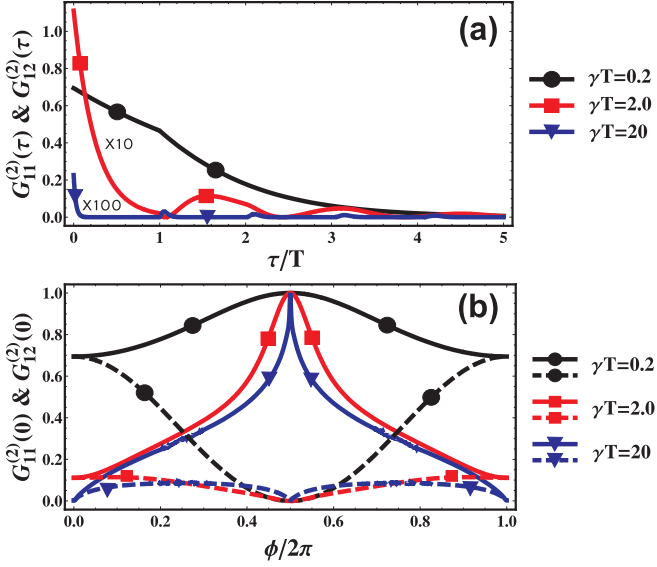}
\caption{{\bf Transmitted and cross second-order correlation functions.} (a) Correlation functions $G_{11}^{(2)}(\tau)$ = $G_{12}^{(2)}(\tau)$. To be visible, we enlarge the correlation functions by $10$ and $100$ times for $\gamma T=2.0$ and $\gamma T=20$, respectively. (b) Correlation functions at $\tau=0$, $G_{11}^{(2)}(0)$ (solid lines) and $G_{12}^{(2)}(0)$ (dashed lines), as functions of the reduced phase $\phi=\varphi -2 k \pi\ \in [0,2\pi]$. Parameters: $\delta/\omega_0=0$ and $\omega_0 T = 2 k \pi$, $k \in \Z$ for both figures. \label{G12function}}
\end{figure}

In \figref{G12function}, we plot the unrenormalized transmitted correlation function $G_{11}^{(2)}(\tau)\approx\mathcal{T}^2g_{11}^{(2)}(\tau)$ and the unrenormalized cross correlation function $G_{12}^{(2)}(\tau)\approx\mathcal{RT}g_{12}^{(2)}(\tau)$, where we only keep the leading order in \eqref{g1mikhail}. In the special case $\delta=0$ and $\omega_0 T = 2 k \pi,\ k \in \Z$, we have $G_{11}^{(2)}(\tau)=G_{12}^{(2)}(\tau)=\frac{1}{4}\abssq{  I_0 (\tau) + I_1 (\tau) }$ as plotted in  \figref{G12function}(a). The equality of $G_{11}^{(2)}(\tau)$ and $G_{12}^{(2)}(\tau)$ can be understood from \eqref{s2final}, which indicates that there is no contribution to the transmitted channel from the first line on the RHS of \eqref{s2final}. Therefore, the second and third lines on the RHS of \eqref{s2final} give equal contribution to the correlation functions $G_{11}^{(2)}(\tau)$ and $G_{12}^{(2)}(\tau)$, respectively. Another feature revealed by \figref{G12function}(a) is that $G_{11}^{(2)}(\tau)$ and $G_{12}^{(2)}(\tau)$ always show bunching behaviours initially, irrespective of the size of the atom, which comes from the fact that phonons are always created in pairs and there is no contribution to the transmitted channel from singe phonon scattering. If we change the phase $\varphi=2 k \pi+\phi$, $\phi\in [0,2\pi]$ and $k \in \Z$, however, the contributions to the correlation functions $G_{11}^{(2)}(\tau)$ and $G_{12}^{(2)}(\tau)$, from the first line on the RHS of \eqref{s2final}, are no longer equal to each other. In \figref{G12function}(b), we plot the initial value of the correlation functions, \textit{i.e.}, $G_{11}^{(2)}(0)$ (solid lines) and $G_{12}^{(2)}(0)$ (dashed lines) as functions of phase $\phi$ for $\gamma T=0.2$ (black, labelled by solid circle), $\gamma T=2.0$ (red, labelled by solid square) and $\gamma T=20$ (blue, labelled by solid triangle). We see that the transmitted correlation function $G_{11}^{(2)}(0)$ is always larger than the cross-correlation function $G_{12}^{(2)}(0)$. In particular, for $\phi = \pi$ we have we have $G_{11}^{(2)}(0)=1$ and $G_{12}^{(2)}(0)=0$, which means that all the phonons are perfectly transmitted and no phonon is reflected back.

\section{\alg{Transient} dynamics with arbitrary drive strength}
\label{sect:cascade}

The finite time delay makes the dynamics of the giant atom highly non-Markovian and with more than a few phonons present in the delay-loop the dynamics is correspondingly complex. Recently, a numerically exact method for integrating the dynamics of open quantum systems with deterministic time-delays was introduced in Ref.~\cite{Grimsmo2015}. The method is based on mapping the problem onto a Markovian problem in an extended system space: It was shown that the problem can be solved by integrating the dynamics of a fictitious quantum cascade \cite{Gardiner1993, Carmichael1993} of system copies, where each copy represents a past version of the atom. This is analogous to how classical stochastic dynamical systems with finite delays can be solved by recasting them in terms of multivariate Markov processes \cite{Frank2002, Whalen2016}. In the following, we use this method to study the atom's dynamics, including properties of the scattered output field, for arbitrary drive strengths.

As discussed in \secref{sec:TheModel}, the giant atom couples to two fields---a left-propagating and a right-propagating---each of which couples to the atom's legs at two different locations, $x = -L/2$ and $x = L/2$. The fields can be treated as independent (correlations between the left- and right-propagating phonons only arise through scattering via the atom), and the atom can thus be seen as being subject to two independent coherent feedback loops \cite{Grimsmo2015}, each with the same time-delay $T = L/v_g$.

We note that only a single feedback field was considered in Ref.~\cite{Grimsmo2015}, but the extension to multiple fields with commensurate delays is straightforward \cite{Whalen2016}. The case we consider here, with two feedback fields with identical delays and a decay rate of $\gamma/2$ into each feedback loop, is particularly straightforward, as from the atom's point of view this is no different than a single feedback loop with a decay rate of $\gamma$. Experimentally, there is of course a difference since the atom has two distinct input-output ports through which a scattered signal can be measured.

In this section, we will explore the giant-atom dynamics beyond the few-phonon limit by considering a monochromatic coherent drive of arbitrary strength applied to the atom. Experimentally this can be achieved by driving the atom through the phonon waveguide, as has been explored for one and two phonons in the previous sections, but one can also consider a drive applied directly on the atom through a voltage side gate. The latter option is more flexible in the sense that a drive with arbitrary frequency can be applied, while a drive with a $\pi$ phase shift between the two legs would cancel if applied through the phonon waveguide.

In either case, the drive can be accounted for by including a drive term in the atom's Hamiltonian
(see \appref{app:inout} for more details)
\be
H_S = \hbar \delta \ket{e}\bra{e} + \frac{1}{2}\hbar \left(\Omega  \sm + \text{H.c.} \right),
\ee
where we have moved to a rotating frame at the drive frequency, $\delta = \omega_d - \omega_0$ is the detuning and $\Omega$ is the drive strength (Rabi frequency). To connect with the treatment in the previous sections, if a drive is applied through the phonon waveguide we have that $\Omega = 2V A^*\left(1 + e^{-i\varphi}\right)$ where $\varphi = \omega_d T$ is the phase shift of the drive, and the coherent input drive is $\alpha_{A}^{\rm in}(t) = A e^{-i\omega_d t}$ as before. Below we however take $\Omega$ to be independent of $\varphi$ since the drive can be applied through a voltage side gate as already mentioned.

Following Ref.~\cite{Grimsmo2015}, to find the atomic state at a time $(k-1)T \le t < kT$ with $ k\in\Z^+$, we numerically solve the cascaded master equation
\be
\frac{\dd}{\dd s} \mathcal{E}_s(t) = \sum_{l = 0}^k \left\{ {-}\frac{i}{2\hbar} \mathcal{H}\left[ H_{l, l+1}(s)\right] + \mathcal{D}\left[ L_{l, l+1}(s)\right] \right\} \mathcal{E}_s(t),
\label{eq:cascade}
\ee
for the atomic time-propagator $\mathcal{E}_s(t)$. This time-propagator is a superoperator on a $k$-fold system space $S_1 \otimes \dots \otimes S_k$, as are the superoperators
\bea
\mathcal{H}[{A}]\sdot &=& [{A},\sdot], \\
\mathcal{D}[{A}]\sdot &=& {A}\sdot {A}^\dag - \half {A}^\dag {A}\sdot - \half \sdot {A}^\dag {A}.
\eea
The system operators ${H}_{l, l+1}$ and ${L}_{l, l+1}$ are given by
\bea
\label{eq:cascops1}
H_{l, l+1} &=& H_S^{(l)} + H_S^{(l+1)} + i \gamma( e^{-i\varphi} \sm^{(l)\dag} \sm^{(l+1)} - \text{H.c.}),\\
L_{l, l+1} &=& \sqrt{\gamma}\sigma_-^{(l)} + \sqrt{\gamma} e^{-i \varphi} \sm^{(l+1)},
\eea
except for $H_{0,1} = H_S^{(1)}$, $H_{k,k+1} = H_S^{(k)}$, $L_{0,1} = \sqrt{\gamma} e^{-i \varphi} \sm^{(1)}$ and $L_{k,k+1} = \sqrt{\gamma} \sm^{(k)}$, where we use a superscript to denote the system on which an operator acts. Finally, we have defined $A^{(l)}(s) = A^{(l)}$ for all $l < k$, and $A^{(k)}(s) = \Theta \left[t - (k - 1)T - s\right] A^{(k)}$, where $\Theta(s)$ is the Heaviside step function, for any system operator $A$.

\begin{figure}
\centering
\includegraphics[width=\linewidth]{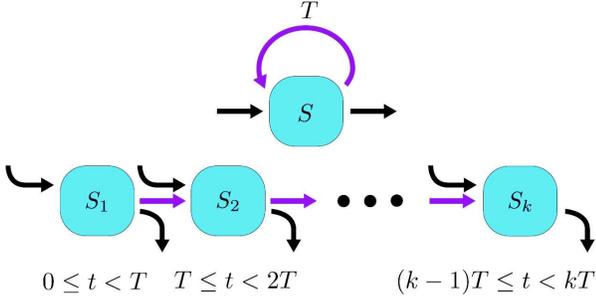}
\caption{Mapping from a quantum system with deterministic time-delay (top) to the fictitious quantum cascade of $k$ identical system copies, described by \eqref{eq:cascade} (bottom). The $l$th copy represents the interval $(l-1)T \le t < l T$. \label{fig:cascade}}
\end{figure}

The cascaded chain given by \eqref{eq:cascade} is illustrated in \figref{fig:cascade}. The mapping from a single system with feedback to a cascaded chain is analogous to the ``method of steps'' used to solve classical delay-differential equations \cite{Frank2002}: the $l$th system copy in the cascade can be interpreted as representing the time-interval $(l-1)T \le t < l T$. A system with feedback is, however, not equivalent to a conventional quantum cascade, since the identical copies do not represent physically distinct systems. This has to be taken into account when the {\it true} reduced density matrix for the system, $\rho_S(t)$, is found by tracing out the auxiliary degrees of freedom. As explained in more detail in Ref.~\cite{Grimsmo2015}, the reduced density matrix is found by first integrating \eqref{eq:cascade} up to $s = T$, where $s$ is an auxiliary time variable, to find $\mathcal{E}_T(t)$ and then acting on the given initial state $\rho_{S_1}(0)$ for system $S_1$ and taking a generalized partial trace:
\be
\rho_S(t) = \gentr_{(S_{k},S_{k-1})} \dots \gentr_{(S_2,S_1)} \, \mathcal{E}_T(t) \rho_{S_1}(0),
\label{eq:rho_S}
\ee
where the generalized trace $\gentr_{(S_{l'},S_{l})}$ acts on a superoperator in the following way
\be
\gentr_{(S_{l'},S_{l})}\,\mathcal{A}\sdot = \sum_{ij} \brakket{i_l}{\mathcal{A}\left(\sdot\otimes\ket{i_{l'}}\bra{j_{l'}}\right)}{j_l},
\ee
where $\ket{i_l}$ and $\ket{i_{l'}}$ are orthonormal bases for the two respective systems, $S_l$ and $S_{l'}$. This operation can be understood as mapping the output of system $S_l$ to the input of system $S_{l'}$ \cite{Grimsmo2015}.

\begin{figure}
\centering
\includegraphics[width=\linewidth]{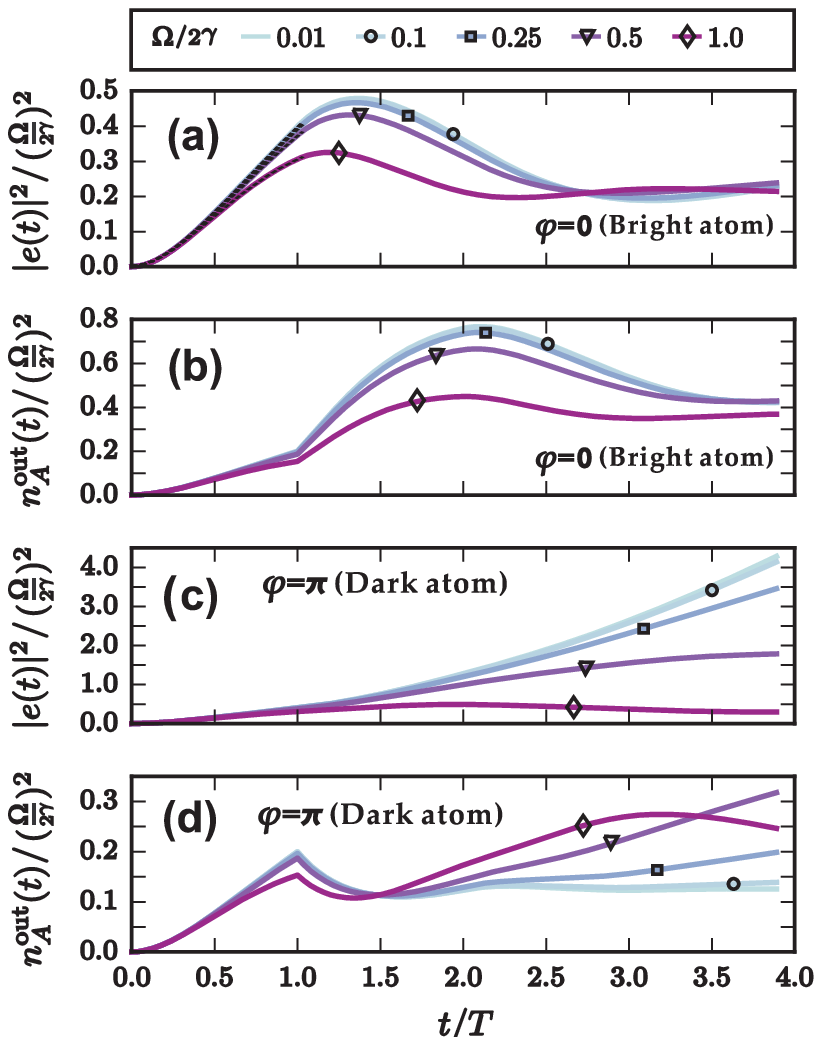}
\caption{\textbf{Transient dynamics for a giant atom starting in the ground state with $\gamma T = 1.0$ (region $C$).} Solid lines show numerically exact results for various drive strengths $\Omega/(2\gamma)$ in the range $0.01$ to $1.0$. The plotted quantities $|e(t)|^2$ and $n^{out}_A(t)$ (normalized by $(\Omega/2\gamma)^2$) are the atomic excited-state probability and outgoing phonon number at leg $A$, respectively. Black dashed lines shown in (a) are the analytical results from Eq.~(\ref{Pe}) for $0<t<T$. Parameters: $\varphi = 0$ (bright atom) for $(a)$ and $(b)$, $\varphi = \pi$ (dark atom) for $(c)$ and $(d)$.\label{fig:shorttime}}
\end{figure}
\begin{figure}
\centering
\includegraphics[width=\linewidth]{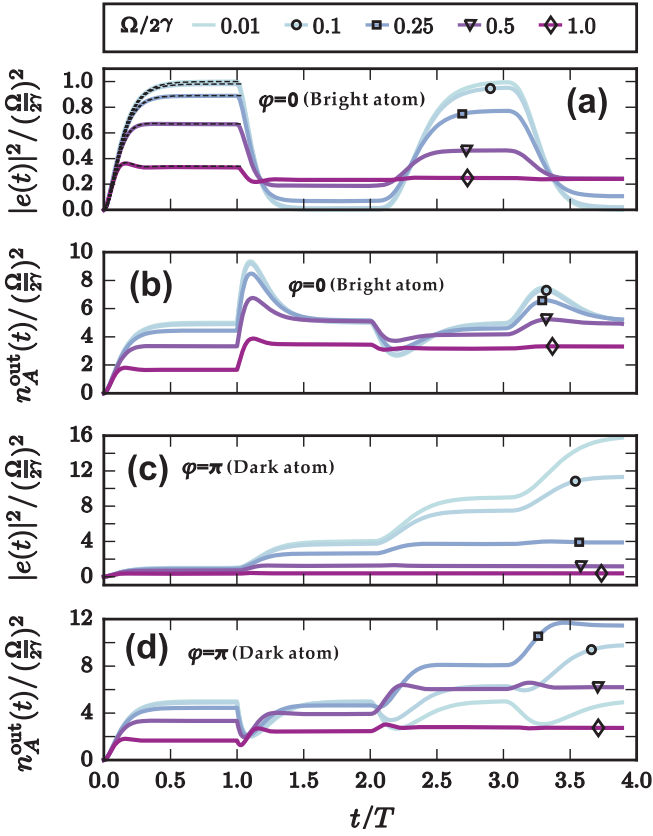}
\caption{\textbf{Transient dynamics for a giant atom starting in the ground state with $\gamma T = 10$ (region $D$).} Solid lines show numerically exact results for various drive strengths $\Omega/(2\gamma)$ in the range $0.01$ to $1.0$.
 The plotted quantities $|e(t)|^2$ and $n^{out}_A(t)$ (normalized by $(\Omega/2\gamma)^2$) are the atomic excited-state probability and outgoing phonon number at leg $A$, respectively. Black dashed lines shown in (a) are the analytical results from Eq.~(\ref{Pe}) for $0<t<T$. Parameters: $\varphi = 0$ (bright atom) for $(a)$ and $(b)$, $\varphi = \pi$ (dark atom) for $(c)$ and $(d)$. Note that the outgoing phonon numbers are larger than that in Fig.~\ref{fig:shorttime} due to the drive being an order of magnitude larger.\label{fig:shorttime2}}
\end{figure}

This method is a powerful tool for exploring the transition from essentially linear dynamics in the single-phonon regime to strongly nonlinear dynamics with multiple phonons. We focus in the following on the giant atom's transient dynamics and the field it emits into the phonon waveguide when the atom is starting in the ground state and driven on resonance with varying drive strengths. The emitted radiation is a phononic analog to resonance fluorescence. To explore the consequences of non-Markovian effects due to the propagation delay between the two legs we consider two values of the time-delay, $\gamma T = 1.0$ and $\gamma T = 10$, corresponding to regions $C$ and $D$ in \figref{fig:ParameterRegimes}, respectively.

In Figs.~\ref{fig:shorttime} $(\gamma T = 1.0)$ and \ref{fig:shorttime2} ($\gamma T = 10$), we display the transient atomic dynamics, starting from the ground state for various drive strengths and an on-resonant drive, $\omega_d = \omega_0$. We plot the short-time evolutions of the atomic excited-state probability $|e(t)|^2$ and the output phonon number at leg $A$, $n_A^{\rm out}(t) \equiv \expec{a_{ A}^{\rm out\ \dag}(t) a_{A}^{\rm out}(t)}$ where the output field $a_A^{\rm out}(t)$ is defined in \appref{app:inout}. We normalize our data by $(\Omega/2\gamma)^2$. We consider two distinct cases with phase shifts of $\varphi = 0$ (bright atom) and $\varphi=\pi$ (dark atom), respectively.

The figures show a clear transition from a linear regime to a nonlinear regime. For low drive strengths, the atomic population as well as the output-field phonon number is proportional to the drive power: the lines with $\Omega/(2\gamma) = 0.01$ and $0.1$ coincide. As the drive power is increased  we enter the nonlinear regime where the two-level nature of the atom starts to be important. The transient dynamics is initially essentially Markovian for $t<T$, as the atom does not feel any feedback effects. Therefore, the atom can be viewed as an ordinary atom with two connections to the waveguide. The probability of excited state has an analytical result \cite{Milburn2008}
\be
|e(t)|^2=\frac{\Omega^2}{(2\gamma)^2 + 2\Omega^2}\left[1 - e^{-\frac{3\gamma t}{2}}\left(\cosh \zeta t + \frac{3\gamma}{2\kappa}\sinh \zeta t \right)\right],
\label{Pe}
\ee
where $\zeta=\sqrt{(\gamma/2)-\Omega^2}$. In Figs.~\ref{fig:shorttime}(a) and \ref{fig:shorttime2}(a), we plot the curve (\ref{Pe}) and compare it with numerical simulation for $0<t<T$. Actually, the time evolutions of $|e(t)|^2$ and $n_A^{out}(t)$ of the bright and dark atoms for $0<t<T$ are the same regardless of the phase $\varphi$ across the two legs. At $t=T$, the atom enters the non-Markovian regime, marked by a sharp kink in the observables, best visible in the output-field phonon number.

In the moderately non-Markovian regime (Fig.~\ref{fig:shorttime}), we see that the feedback interferes constructively with the outfield for the bright atom (Fig.~\ref{fig:shorttime}(b)). For the dark atom (Fig.~\ref{fig:shorttime}(d)), the destructive interferences instead leads to a reduction of the output field. The bright atom approaches a steady-state within the simulation time, due to the comparatively strong dissipation. The dark atom has an effectively weaker coupling to the outgoing phonons, leading to an increase of the atom population during the whole simulation for all but the strongest drive. For the strongest drive, the atom population changes noticeably during the propagation time $T$, which makes the destructive interference of the feedback less efficient.

In the deep non-Markovian regime (Fig.~\ref{fig:shorttime2}), we find transient dynamics with plateaus of constant population and output field amplitude, interrupted approximately at integer values of $t/T$ with steps on the timescale of $1/\gamma$. This pattern can be understood starting from the initial Markovian regime $t<T$, where the steady state of the driven, damped atom is established on the timescale of the local coupling strength $1/\gamma$. At $t=T$ the feedback changes the effective drive and damping, which gives a transient period until a new steady state is established. The length of the transient periods increase with each period, probably allowing the system to approach a global steady state at very long times.
For the bright atom (Fig.~\ref{fig:shorttime2}(a)), we note a pattern of alternating high and low population where the relative amplitude of the steps decrease with increasing drive strength and increasing time. For the dark atom (Fig.~\ref{fig:shorttime2}(c)), we instead see a population increasing with time, again due to destructive interference in the output fields.

\begin{figure}
\centering
\includegraphics[width=\linewidth]{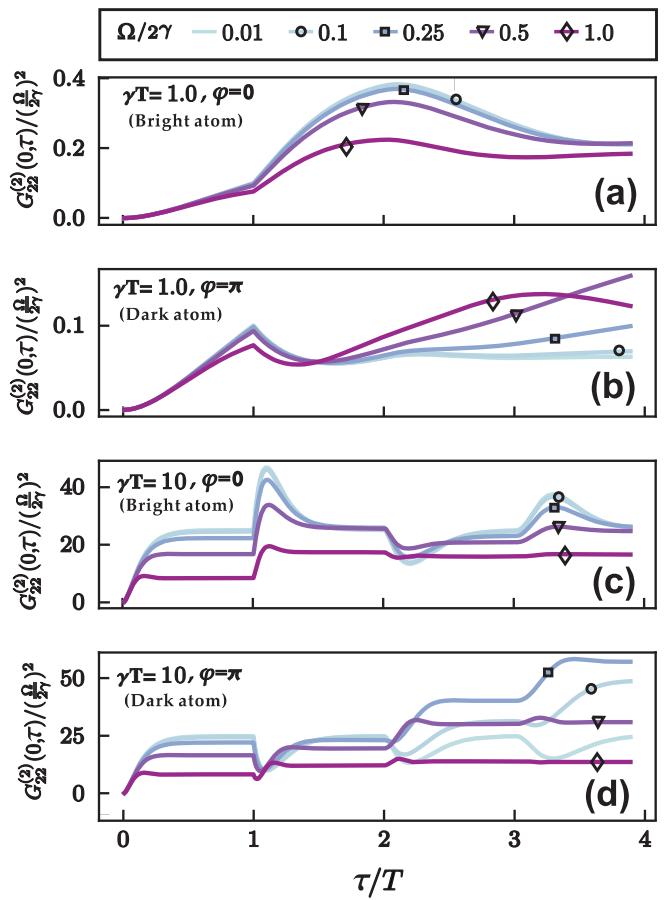}
\caption{\textbf{ Phonon correlation function $G^{(2)}_{22}(0,\tau)$ scaled by $(\Omega/2\gamma)^2$ for a giant atom starting in the excited state.} (a) $\gamma T = 1.0$, $\varphi = 0$ (bright atom), (b) $\gamma T = 1.0$, $\varphi = \pi$ (dark atom), (c) $\gamma T = 10.0$, $\varphi = 0$ (bright atom), (d) $\gamma T = 10.0$, $\varphi = \pi$ (dark atom). \label{fig:shorttime_g2}}
\end{figure}

It is also interesting to look at higher-order correlation functions for the atom's output field. In \figref{fig:shorttime_g2} we show the second-order correlation function
\be
G^{(2)}_{22}(t_0,\tau) \equiv \expec{a_{ A}^{\rm out \ \dag}(t_0) a_{A}^{\rm out \ \dag}(t_0 + \tau) a_{A}^{\rm out}(t_0 + \tau) a_{A}^{\rm out}(t_0)}.
\label{g2t0t}
\ee
In our previous definition of second-order correlation function in \eqref{G2Mikhal}, we have implicitly chosen the average over the stationary state after scattering $\ket{\Psi_f}$, \textit{i.e.}, we calculated the correlation function in \eqref{g2t0t} at $t_0 = +\infty$. Since $G^{(2)}_{22}(t_0,\tau) = 0$ for an atom starting in the ground state, we show in \figref{fig:shorttime_g2} the behaviour when starting from the excited state instead. We again show results for two values of the time-delay, $\gamma T = 1.0$ and $\gamma T = 10.0$, as well as two values of the phase shift $\varphi = 0$ (bright atom) and $\varphi = \pi$ (dark atom). From the definition of the second-order correlation function, $G^{(2)}_{22}(0,\tau)$ is proportional to the joint probability density of observing one phonon at $t=0$ and another at $t=\tau$ \cite{Scully1997}. We have assumed that the atom is in the excited state and the entire waveguide (including the part between the atom's legs) is in the vacuum state at initial time $t=0$. Thus the atom is in the ground state when the first phonon is observed at $t=0$. Then the probability to observe another phonon at $t=\tau$ is proportional to the phonon field emitted by the atom at leg $A$. Therefore, the second-order correlation functions $G^{(2)}_{22}(0,\tau)$ scaled by $(\Omega/2\gamma)^2$ in Fig.~\ref{fig:shorttime_g2} exhibit exactly the same behaviors of $n_A^{out}(t)$, up to a normalization factor, as shown in Figs.~\ref{fig:shorttime} and \ref{fig:shorttime2}. The second-order correlation function shows particularly strong signatures of the feedback force at times $\tau = nT$ for integer $n$.

Output field properties are calculated using a generalization of the well-known quantum regression formula for systems with time-delays, given in \appref{app:outfield}. The numerical simulations were performed using an open source implementation of the method from Ref.~\cite{Grimsmo2015} in QuTiP \cite{Johansson2012, Johansson2013}.

\section{Summary and outlook}

We have investigated the quantum dynamics of a single two-level quantum system (a transmon qubit) coupled to a SAW transmission line via two connection points  {separated by a large distance $L$, which introduces a deterministic time delay $T$.  We explored how the non-Markovian dynamics that arises due to a long time delay affects the spontaneous emission and scattering properties of this system, which we call a giant atom. We found several notable differences to the more common case of a small atom coupled to a transmission line at a single point. Both the large time delay and the phase acquired by phonons travelling between the connection points, resulting in interference effects, are important to explain these differences.

For single-phonon processes, we obtained analytical solutions by solving a differential time-delay equation. Using these solutions, we first studied the power spectrum of the spontaneous emission from the giant atom. This revealed the presence of several frequency modes, something not seen for a small atom. Furthermore, interference between these modes was shown to make the energy decay from the system polynomial at first; only after a long time, when all modes but one has decayed, does the giant atom follow an exponential decay law. During this process, the atom experiences revivals as it emits energy from one connection point and later reabsorbs some of it at the other one. The presence of multiple modes at large $T$ was also shown to cause multiple peaks in the single-phonon reflection of the giant atom, another feature distinguishing it from a small atom, which only has a single reflection peak at its resonance frequency.

For two-phonon processes, we obtained an exact analytical solution of the scattering matrix by using the diagrammatic Lippmann-Schwinger-equation approach given in Ref.~\cite{Laakso2014}. Using the two-phonon scattering matrix, we calculated the lowest-order correction to the transmittance of the system. For a small atom, increasing the driving always increases the transmittance, but for the case of a giant atom, we show that the transmittance sometimes decreases instead. This is due to interference effects between phonons emitted at the two different connection points. We also calculated the inelastic (incoherent) power spectrum. For a small atom ($\gamma T<1$), the inelastic power spectrum showed a central peak around the driving frequency. For $\gamma T>1$, the central peak splits into two peaks due to the time delay of the giant atom. This is different from the Mollow triplet, which is due to strong driving.

We discussed the second-order correlation functions for phonons scattered by the giant atom. While phonons (or photons) reflected from a small atom will display perfect antibunching, this effect is diminished for a giant atom since a second phonon can be emitted from the second connection point at an earlier time. However, the second-order correlation function for the reflected phonons from a giant atom has a richer structure than for the case of a small atom; both bunching and antibunching occur, and the function has kinks at integer multiples of $T$.

Finally, we also considered coherent driving of arbitrary strength being applied to the giant atom. In this case, an analytical solution is beyond the reach of the diagrammatic approach. Therefore, we instead used the exact numerical method for integrating the dynamics of open quantum systems with deterministic time delays introduced in Ref.~\cite{Grimsmo2015}. This allowed us to numerically simulate the short-time dynamics of the giant atom and calculate second-order correlation functions.

There are several possible directions of research beyond our present work. When it comes to an experimental implementation, we believe that the parameter regimes we have considered here are rather straightforward to reach with a transmon coupled to SAWs by modifying the experimental setup of Ref.~\cite{Gustafsson2014}. A pure circuit-QED setup might also be able to reach a regime with long enough time delays to demonstrate differences from the small-atom case, but to achieve truly long time delays SAWs seem more promising. One potential obstacle for measurements in such experiments is the low conversion efficiency of SAWs to electric microwave signals in conventional symmetric IDTs. However, the recent work in Ref.~\cite{Delsing2016} demonstrated unidirectional transducers (UDTs), which can increase the conversion efficiency up to $99.4\%$ at GHz frequencies and millikelvin temperatures. From a technical perspective, it is of interest whether the diagrammatic Lippmann-Schwinger approach can be extended to scattering with more than two phonons. In a similar vein, it would be desirable to extend the numerical technique used to simulate the short-time dynamics to work for longer times. Finally, the system under investigation could also be extended, \textit{e.g.}, to include more than two connection points of the atom or to include several giant atoms coupled to the SAW transmission line.

\bigskip
\textbf{Acknowledgements}
We acknowledge helpful discussions with Prof.~Alexandre Blais.
L. G. acknowledges financial support from Carl-Zeiss Stiftung (0563-2.8/508/2). A.F.K. acknowledges support from a JSPS Postdoctoral Fellowship for Overseas Researchers.
G. J. acknowledges financial support from the Swedish Research Council and the Knut and Alice Wallenberg Foundation.
A.L.G. is supported by NSERC. This research was undertaken thanks in part to funding from the Canada First Research Excellence Fund.

\appendix

\section{Single-phonon processes}
\label{app:SinglePhonon}

\subsection{Equations of motion}
\label{app:SinglePhononEOM}

We start from the RWA Hamiltonian based on \eqref{Hv}
\bea
&&H = -\frac{1 + \sigma_z}{2} \delta + \sum_{\alpha = 1, 2} \int d\omega \omega a^\dag_{\alpha \omega} a_{\alpha \omega} \nn\\
&&+ \sqrt{\frac{\gamma}{4\pi}} \int d\omega \left\{ \sm \left[ a^\dag_{1 \omega} e^{-i (\omega T + \varphi)/2} + a^\dag_{2 \omega} e^{i (\omega T + \varphi)/2} \right] + \mathrm{H.c.} \right\} \nn\\
&&+ \sqrt{\frac{\gamma}{4\pi}} \int d\omega \left\{ \sm \left[ a^\dag_{1 \omega} e^{i (\omega T + \varphi)/2} + a^\dag_{2 \omega} e^{-i (\omega T + \varphi)/2} \right] + \mathrm{H.c.} \right\}, \nn\\
{}
\eea
where $\delta \equiv \omega_d - \omega_0$ is the atomic detuning and $\sz \equiv \ketbra{e}{e} - \ketbra{g}{g}$. The phonon frequency $\omega \equiv \omega_p - \omega_d$ is also shifted by the rotating frequency. The phase difference between two legs is given by $\varphi = \omega_d T = \omega_d L/v_g$.

As discussed in \secref{sec:TheModel}, we make the following ansatz for the single-phonon process
\be
\ket{\Psi(t)} = \int d\omega \left[ \alpha_{1 \omega}(t) a^\dag_{1 \omega} + \alpha_{2 \omega}(t) a^\dag_{2 \omega} \right] \ket{g,vac} + e(t) \ket{e,vac}.
\ee
Then, the Schr\"odinger equation gives
\bea
&&H \ket{\Psi(t)} = - \delta e(t) \ket{e,vac} \nn\\
&&+ \int d\omega \left[ \omega \alpha_{1 \omega}(t) a^\dag_{1 \omega} + \omega \alpha_{2 \omega}(t) a^\dag_{2 \omega}\right] \ket{g,vac} \nn\\
&&+ e(t) \sqrt{\frac{\gamma}{4\pi}} \int d\omega \left[a^\dag_{1 \omega} e^{-i (\omega T + \varphi)/2} + a^\dag_{2 \omega} e^{i (\omega T + \varphi)/2} \right] \ket{g,vac} \nn\\
&&+ e(t) \sqrt{\frac{\gamma}{4\pi}} \int d\omega \left[ a^\dag_{1 \omega} e^{i (\omega T + \varphi)/2} + a^\dag_{2 \omega} e^{-i (\omega T + \varphi)/2} \right] \ket{g,vac} \nn\\
&&+ \sqrt{\frac{\gamma}{4\pi}} \int d\omega \left[ \alpha_{1 \omega} e^{i (\omega T + \varphi)/2} + \alpha_{2 \omega} e^{-i (\omega T + \varphi)/2} \right] \ket{e,vac} \nn\\
&&+ \sqrt{\frac{\gamma}{4\pi}} \int d\omega \left[ \alpha_{1 \omega} e^{-i (\omega T + \varphi)/2} + \alpha_{2 \omega} e^{i (\omega T + \varphi)/2} \right] \ket{e,vac} \nn\\
&& = i \int d\omega \left[ \dot{\alpha}_{1 \omega}(t) a^\dag_{1 \omega} + \dot{\alpha}_{2 \omega}(t) a^\dag_{2 \omega} \right] \ket{g,vac} + i \dot{e}(t) \ket{e,vac}. \qquad
\eea
Therefore, we have the dynamical equations for the giant atom
\bea
&&\frac{d}{dt}e(t) = i\delta e(t) \nn\\
&&-i \sqrt{\frac{\gamma}{4\pi}} \int d\omega \left[ e^{i (\omega T + \varphi)/2} + e^{-i (\omega T + \varphi)/2} \right] \alpha_{1 \omega}(t) \nn\\
&&-i \sqrt{\frac{\gamma}{4\pi}} \int d\omega \left[ e^{-i (\omega T + \varphi)/2} + e^{i (\omega T + \varphi)/2} \right] \alpha_{2 \omega}(t),
\label{atomEq}
\eea
right-propagating phonon fields in the transmission line
\be
\frac{d}{dt}\alpha_{1 \omega}(t) = -i \omega \alpha_{1 \omega}(t) -i e(t) \sqrt{\frac{\gamma}{4\pi}} \left[e^{-i (\omega T + \varphi)/2}
+ e^{i (\omega T + \varphi)/2} \right],
\label{rightEq}
\ee
and left-propagating phonon fields in the transmission line
\be
\frac{d}{dt}\alpha_{2 \omega}(t) = -i \omega \alpha_{2 \omega}(t) - i e(t) \sqrt{\frac{\gamma}{4\pi}} \left[e^{i (\omega T + \varphi)/2} + e^{-i (\omega T + \varphi)/2} \right].
\label{leftEq}
\ee
Integrating Eqs.~(\ref{rightEq}) and (\ref{leftEq}), we have
\bea
&&\alpha_{1 \omega}(t) = e^{-i \omega t}\Big\{ \alpha_{1 \omega}(0)  \nn\\
&&  -i \sqrt{\frac{\gamma}{4\pi}} \left[e^{-i (\omega T + \varphi)/2} + e^{i (\omega T + \varphi)/2} \right] \int_0^t dt' e(t') e^{i \omega
t'} \Big\}
\label{rightS}
\eea
and
\bea
&&\alpha_{2 \omega}(t) = e^{-i \omega t} \Big\{ \alpha_{2 \omega}(0)  \nn\\
&&  -i \sqrt{\frac{\gamma}{4\pi}} \left[ e^{i (\omega T + \varphi)/2} + e^{-i (\omega T + \varphi)/2} \right] \int_0^t dt' e(t') e^{i \omega t'} \Big\}.
\label{leftS}
\eea
Inserting the two equations above into \eqref{atomEq}, we have
\bea
&& \frac{d}{dt}e(t) = i \delta e(t) \nn\\
&& -i \sqrt{\frac{\gamma}{4\pi}} \int d\omega \left[e^{i (\omega T + \varphi)/2} + e^{-i (\omega T + \varphi)/2} \right] e^{-i \omega t}\alpha_{1 \omega}(0) \nn\\
&& -i \sqrt{\frac{\gamma}{4\pi}} \int d\omega \left[e^{-i (\omega T + \varphi)/2} + e^{i (\omega T + \varphi)/2} \right] e^{-i \omega t}\alpha_{2 \omega}(0) \nn\\
&& -\frac{\gamma}{2\pi} \int d\omega \abssq{e^{i (\omega T + \varphi)/2} + e^{-i (\omega T + \varphi)/2}} \int_0^t dt' e(t') e^{-i \omega
(t-t')}. \quad\quad
\label{atomS}
\eea
If the amplitude of one plane wave at time $t$ is $\alpha_{1(2) \omega}(t)$, then the total SAW field in the transmission line at position $x$ is
\be
\alpha_{1(2)}(x,t) \equiv \frac{1}{\sqrt{2 \pi v_g}} \int d\omega e^{\pm i \omega x/v_g} \alpha_{1(2) \omega}(t).
\label{alphaxt1}
\ee
For right- (left-) propagating fields we take positive (negative) sign in the phase. Using this notation, we can continue {massaging} \eqref{atomS}:
\bea
&& \frac{d}{dt}e(t) = i \delta e(t) \nn\\
&& -i \sqrt{\frac{\gamma v_g}{2}} \left[ e^{i \varphi/2} \alpha_1(L/2 - v_g t, 0) + e^{-i \varphi/2} \alpha_1(-L/2 - v_g t, 0) \right] \nn\\
&& -i \sqrt{\frac{\gamma v_g}{2}} \left[ e^{-i \varphi/2} \alpha_2(L/2 - v_g t, 0) + e^{i \varphi/2} \alpha_2(-L/2 - v_g t, 0) \right] \nn\\
&& -\gamma \int_0^t dt' e(t') \left[ 2 \delta(t - t') + e^{i \varphi} \delta(T - t + t') \right. \nn\\
&& \left. + e^{-i \varphi} \delta(-T - t + t') \right] \nn\\
&& = i \delta e(t) - \gamma \left[ e(t) - \Theta(t - T) e^{i \varphi} e(t - T) \right] \nn\\
&& - i \sqrt{\frac{\gamma v_g}{2}} \left[ e^{i \varphi/2} \alpha_1(L/2 - v_g t, 0) + e^{-i \varphi/2} \alpha_1(-L/2 - v_g t, 0) \right] \nn\\
&& - i \sqrt{\frac{\gamma v_g}{2}} \left[ e^{-i \varphi/2} \alpha_2(L/2 - v_g t, 0) + e^{i \varphi/2} \alpha_2(-L/2 - v_g t, 0) \right], \nn\\
{}
\label{atomS1}
\end{eqnarray}
where $\Theta(x)$ is the Heaviside step function ($\Theta(x) = 0$ for $x < 0$ and $\Theta(x) = 1$ for $x > 0$). We can transform \eqref{atomS1} into the original frame by making the replacement $\tilde{e}(t) \to e^{-i \omega_d t} e(t)$. We then get
\bea
\frac{d}{dt}\tilde{e}(t) &=& -i \omega_0 \tilde{e}(t) - \gamma \left[ \tilde{e}(t) - \Theta(t - T) \tilde{e}(t - T) \right] \nn\\
&& - i V \left[ \tilde{\alpha}_1(L/2 - v_g t, 0) + \tilde{\alpha}_1(-L/2 - v_g t, 0) \right. \nn\\
&& \left. + \tilde{\alpha}_2(L/2 - v_g t, 0) + \tilde{\alpha}_2(-L/2 - v_g t, 0) \right],
\eea
where $V = \sqrt{\frac{\gamma v_g}{2}}$ is the coupling strength and
\be
\tilde{\alpha}_{1(2)}(x,0) \equiv e^{ \pm i \omega_d x/v_g} \alpha_{1(2)}(x, 0)
\label{atomS2}
\ee
are the SAW fields in the rest frame. For simplicity in the following, we further omit the tildes. Then, we have
\bea
\frac{\partial e(t)}{\partial t} &=& - i \omega_0 e(t) - \gamma \left[ e(t) + e(t - T) \right] \nn\\
&& - i V \left[ \alpha_1(L/2 - v_g t, 0) + \alpha_1(-L/2 - v_g t, 0) \right. \nn\\
&& \left. + \alpha_2(L/2 - v_g t, 0) + \alpha_2(-L/2 - v_g t, 0) \right].
\label{EOMApp}
\end{eqnarray}
Here, we have assumed $e(t)=0$ for $t<0$ and thus neglected the Heaviside step function $\Theta(t-T)$.  By Fourier transforming \eqref{EOMApp}, \textit{i.e.},
\bea
e(\omega) &=& \int_{-\infty}^{+\infty} dt e(t) e^{i\omega t}, \quad e(t < 0) = 0 \\
e(t) &=& \frac{1}{2\pi} \int d\omega e(\omega) e^{-i\omega t}, \quad t > 0
\eea
and using \eqref{alphaxt1}, we have
\bea
&& - e(0) - i \omega e(\omega) = - i \omega_0 e(\omega) - \gamma e(\omega) - \gamma e^{i \omega T} e(\omega) \nn\\
&& - i V \sqrt{\frac{2\pi}{v_g}} \left[ \alpha_{1 \omega}(0) + \alpha_{2 \omega}(0) \right] \left( e^{i \omega T/2} + e^{-i \omega T/2} \right).
\eea
Therefore, we have the solutions
\be
e(\omega) = \frac{i e(0) + V \sqrt{2\pi/v_g} \left[ \alpha_{1 \omega}(0) + \alpha_{2 \omega}(0) \right] \left( e^{i \omega T/2} + e^{-i \omega T/2} \right)}{\omega - \omega_0 + i \gamma + i \gamma e^{i \omega T}}
\ee
and
\bea
&& e(t) = \frac{i e(0)}{2\pi} \int d\omega \frac{ e^{-i \omega t}}{\omega - \omega_0 + i \gamma + i \gamma e^{i \omega T}} \nn\\
&+& \frac{V}{\sqrt{2\pi v_g}} \int d\omega \frac{\left[ \alpha_{1 \omega}(0) + \alpha_{2 \omega}(0) \right] \left( e^{i \omega T/2} + e^{-i \omega T/2} \right)}{\omega - \omega_0 + i \gamma + i \gamma e^{i \omega T}} e^{-i \omega t}. \nn\\
{}
\label{appet}
\eea

\subsection{Spontaneous emission}
\label{app:SinglePhononSpontaneousEmission}

Assuming the initial condition $e(0) = 1$ for the giant atom and no driving, \textit{i.e.}, $\alpha_{1 \omega}(0) = \alpha_{2 \omega}(0) = 0$, we have from \eqref{appet}
\bea
e(t) &=& \frac{i}{2\pi} \int d\omega \frac{ e^{-i \omega t}}{\omega - \omega_0 + i\gamma + i \gamma e^{i \omega T}} \nn\\
&=& \frac{i}{2\pi} \int d\omega \frac{ e^{-i \omega t}}{\omega - \omega_0 + i \gamma} \left(1 + \frac{i \gamma e^{i \omega T}}{\omega - \omega_0 + i \gamma} \right)^{-1} \nn\\
&=& \frac{i}{2\pi} \int d\omega \frac{ e^{-i \omega t}}{\omega - \omega_0 + i \gamma} \sum_{n = 0}^{+\infty} \left(-\frac{i \gamma e^{i \omega T}}{\omega - \omega_0 + i \gamma} \right)^n \nn\\
&=& \sum_{n = 0}^{+\infty} \frac{i}{2\pi} \int d\omega \frac{(-i \gamma)^n e^{-i \omega
(t - n T)}}{(\omega - \omega_0 + i \gamma)^{n + 1}} \nn\\
&=& \sum_{n = 0}^{+\infty} \Theta(t - n T) \frac{[ - \gamma (t - n T) ]^n}{n!} e^{-i (\omega_0 - i \gamma) (t - n T)}.
\eea
This is in fact equivalent to the solution given as \eqref{et} in the main text. The solution can also be put into the alternative form
\bea
e(t) &=& \frac{i}{2\pi} \int d\omega \frac{ e^{-i \omega t}}{\omega - \omega_0 + i \gamma + i \gamma e^{i \omega T}} \nn\\
&=& \sum_{k} \frac{ e^{-i \omega_{(k)} t}}{1 - \gamma T e^{i \omega_{(k)} T}}.
\eea
Here, we have used the residue theorem and the poles $\omega_{(k)}$ are given by equation
\be
\omega_{(k)} - \omega_0 + i \gamma + i \gamma e^{i \omega_{(k)} T} = 0,
\ee
or the following equivalent form:
\be
\left[ -i (\omega_{(k)} - \omega_0 + i \gamma) T \right] e^{-i (\omega_{(k)} - \omega_0 + i \gamma) T} = - \gamma T e^{\gamma T + i \omega_0 T}.
\ee
The solutions are given by \eqref{LambertW} in the main text, $\textit{i.e.},$
\begin{eqnarray}
\omega_{(k)} = \omega_0 - i \gamma + i \frac{1}{T} W_k \left(-\gamma T e^{\gamma T + i \omega_0 T}\right),
\label{polesapp}
\end{eqnarray}
with $k \in\Z$. Here, $W(z)$ is the Lambert W-function \cite{Corless1996} defined by $z = W(z) e^{W(z)}$, which in general is a multivalued function with branches $W_k(z),\ \in\Z$. The asymptotic behaviour of $W(-re^{i\theta})$ in the limit $r\gg 1$ can be obtained in the following way. Starting from the definition $-r e^{i \theta} = W(-r e^{i \theta}) e^{W(-r e^{i \theta})}$ and $-re^{ i \theta} = r e^{i (2k+1) \pi + i \theta}$ with $\theta \in [0,2\pi)$, we have
\be
W_k = \ln r + i [(2k+1) \pi + \theta] - \ln W_k.
\label{Witerate}
\ee
As the lowest-order approximation we can neglect $\ln W_k$ and have $W_k \approx \ln r + i [(2k+1) \pi + \theta]$. Plugging this solution back into \eqref{Witerate}, we get the higher-order solution
\bea
W(-r) &\approx & \ln \frac{r}{\sqrt{(\ln r)^2 + [(2k + 1) \pi + \theta]^2}} \nn\\
&&+ i \left[ \pi (2k + 1) + \theta - \arctan \frac{\pi (2k + 1) + \theta}{\ln r} \right]. \quad\quad
\label{Wsolution}
\eea
This approximate solution is valid for $r \gg 1$. It can be obtained from Eq.~(4.20) in Ref.~\cite{Corless1996}. We now apply this approximate solution to \eqref{LambertW}. By replacing $r = \gamma T e^{\gamma T}$ and assuming $\omega_0 T = 2n \pi + \Delta \pi$ with $n\in\Z$ and $\Delta \in [0,2)$, we have the frequency modes for $\gamma T e^{\gamma T} \gg 1$
\bea
\omega_{(k)} &\approx& \omega_0 - \frac{1}{T} \left[ \pi (2k + 1 + \Delta) - \arctan \frac{\pi (2k + 1 + \Delta)}{\gamma T} \right] \nn\\
&& - i \frac{1}{2T} \ln \left(1 + \left[ \frac{(2k + 1 + \Delta) \pi}{\gamma T} \right]^2 \right).
\eea
For $\gamma T \gg 1$, the frequency interval is $\Delta \omega \approx 2\pi/T.$

We now study the emission spectrum of outgoing phonons. By defining the variables in the original frame, \textit{i.e.}, $\tilde{\alpha}_{1(2) \omega} \equiv e^{-i \omega_d t} \alpha_{1(2) \omega}(t)$, $\tilde{e}(t') = e^{-i \omega_d t'} e(t')$ and $\tilde{\omega} \equiv \omega + \omega_d$, Eqs.~(\ref{rightS}) and (\ref{leftS}) can be written as
\bea
&&{\alpha}_{1(2) \omega}(t) = e^{-i {\omega} t}{\alpha}_{1(2) \omega}(0) \nn\\
&& -i \sqrt{\frac{\gamma}{4\pi}} \left( e^{-i \omega T/2} + e^{i \omega T/2} \right) \int_0^t dt' e(t') e^{i \omega (t' - t)}.
\label{sawfield}
\eea
As done in \eqref{EOMApp}, we have omitted the tildes for simplicity. Without driving (\textit{i.e.}, $\alpha_{1 \omega}(0) = 0$ and $\alpha_{2 \omega}(0) = 0$), we have $\alpha_\omega^{out}(t) = \alpha_{1 \omega}(t) = \alpha_{2 \omega}(t)$ from symmetry and
\bea
\alpha_\omega^{out}(t)&=& - i \sqrt{\frac{\gamma}{4\pi}} \left( e^{-i \omega T/2} + e^{i \omega T/2} \right) \int_0^t dt' e(t') e^{i \omega (t' - t)} \nn\\
&=& \frac{1}{2\pi} \sqrt{\frac{\gamma}{4\pi}} \left( e^{-i \omega T/2} + e^{i \omega T/2} \right) \nn\\
&\times& \int d\omega' \frac{ e^{-i \omega' t}}{\omega' - \omega_0 + i \gamma + i \gamma e^{i \omega' T}} \int_0^t dt' e^{i (\omega - \omega') (t' - t)} \nn\\
&=& \frac{1}{2\pi} \sqrt{\frac{\gamma}{4\pi}} \left(e^{-i \omega T/2} + e^{i \omega T/2} \right) \nn\\
&\times& \int d\omega' \frac{1}{\omega' - \omega_0 + i \gamma + i \gamma e^{i \omega' T}} \frac{e^{-i \omega' t} - e^{-i \omega t}}{i (\omega - \omega')} \nn\\
&=& - \sqrt{\frac{\gamma}{4\pi}} \left( e^{-i \omega T/2} + e^{i \omega T/2} \right)  \nn\\
&\times& \sum_k \frac{1}{1 - \gamma T e^{i \omega_{(k)} T}} \frac{e^{-i \omega_{(k)} t} - e^{-i \omega t}}{\omega - \omega_{(k)}}.
\eea
Here, we have assumed $e^{i \omega_0 T} \neq -1$. Then all the poles given by \eqref{polesapp} are in the lower half-plane, and we can write
\be
\frac{1}{\omega' - \omega_0 + i \gamma + i \gamma e^{i \omega' T}} = \sum_k \frac{1}{1 - \gamma T e^{i \omega_{(k)} T}} \frac{1}{\omega - \omega_{(k)}}.
\label{ID}
\ee
In the long-time limit, $e^{-i \omega_{(k)} t} \to 0$ due to the negative imaginary part of $\omega_{(k)}$, and therefore
\bea
\alpha_{\omega}^{out}(+\infty) &=& e^{-i \omega t} \sqrt{\frac{\gamma}{\pi}} \cos(\omega T/2) \sum_k \frac{1}{1 - \gamma T e^{i \omega_{(k)} T}} \frac{1}{\omega - \omega_{(k)}} \nn\\
&=& e^{-i \omega t} \sqrt{\frac{\gamma}{\pi}} \frac{\cos(\omega T/2)}{\omega - \omega_0 + i \gamma + i \gamma e^{i \omega T}}.
\eea
We define the emission spectrum of outgoing phonons
\be
S^{out}(\omega) \equiv 2 \pi \abssq{\alpha_{\omega}^{out}(+\infty)} = \frac{\gamma(1 + \cos \omega T)}{\abssq{\omega - \omega_0 + i \gamma + i \gamma e^{i \omega T}}},
\ee
which can be proven to be equivalent to \eqref{OutputSpectrum} in the main text using the identity in \eqref{ID}.

\subsection{Boundary conditions}

Using Eqs.~(\ref{alphaxt1}) and (\ref{sawfield}), we compute the SAW fields in the transmission line
\bea
\alpha_1 (x,t) &=& \alpha_1 (x - v_g t, 0) \nn\\
&& - i \sqrt{\frac{\gamma}{2 v_g}} \left[ \Theta(x - L/2) e(t + T/2 - x/v_g) \right. \nn\\
&& \left. + \Theta(x + L/2) e(t - T/2 - x/v_g) \right],
\label{SAWfield1} \\
\alpha_2 (x, t) &=& \alpha_2 (x - v_g t, 0) \nn\\
&& - i \sqrt{\frac{\gamma}{2 v_g}} \left[ \Theta(-x - L/2) e(t + T/2 + x/v_g) \right. \nn\\
&& \left. + \Theta(-x + L/2) e(t - T/2 + x/v_g) \right].
\label{SAWfield2}
\eea
In particular, we have the boundary conditions at the two legs for the right-propagating SAW field
\bea
\alpha_1 (+L/2 + 0^+, t) &=& \alpha_1 (+L/2 - v_g t, 0) \nn\\
&& - i \sqrt{\frac{\gamma}{2 v_g}} \left[ e(t) + e(t - T) \right], \nn\\
\alpha_1 (-L/2 - 0^-, t) &=& \alpha_1 (-L/2 - v_g t, 0),
\label{connection1}
\eea
and for the left-propagating SAW field
\bea
\alpha_2 (-L/2 - 0^-, t) &=& \alpha_2 (-L/2 - v_g t, 0) \nn\\
&& - i \sqrt{\frac{\gamma}{2 v_g}} \left[ e(t) + e(t - T) \right], \nn\\
\alpha_2 (+L/2 + 0^+, t) &=& \alpha_2 (+L/2 - v_g t, 0).
\label{connection2}
\eea
Introducing the following notation
\bea
\alpha_A^{in}(t) &\equiv& \alpha_1 (-L/2 - 0^-, t) = \alpha_1 (-L/2 - v_g t, 0), \nn\\
\alpha_B^{in}(t) &\equiv& \alpha_2 (+L/2 - 0^+, t) = \alpha_2 (+L/2 - v_g t, 0), \nn\\
\alpha_A^{out} &\equiv& \alpha_2 (-L/2 - 0^+, t), \nn\\
\alpha_B^{out} &\equiv& \alpha_1 (+L/2 + 0^+, t), \nn
\eea
we can rewrite the boundary conditions as
\bea
\alpha_B^{out}(t) &\equiv& \alpha_A^{in}(t - T) - i \sqrt{\frac{\gamma}{2 v_g}} \left[ e(t) + e(t - T) \right], \label{connection3a}\\
\alpha_A^{out}(t) &\equiv& \alpha_B^{in}(t - T) - i \sqrt{\frac{\gamma}{2 v_g}} \left[ e(t) + e(t - T) \right]. \label{connection3}
\eea
With this notation, \eqref{EOMApp} recovers \eqref{EOMb} in the main text.

\subsection{Polynomial decay}
\label{app:SinglePhononPolynomialDecay}

Without driving, the spontaneous emission of the giant atom excites SAW wave-packets in both directions at the two legs. We are interested in the phonon field excited between the two legs, \textit{i.e.}, $-L/2 < x < L/2$. The total field is the superposition of the two fields which can be obtained from Eqs.~(\ref{SAWfield1}) and (\ref{SAWfield2}),
\bea
&& \alpha(x, t) = \alpha_1 \left( -\frac{L}{2} < x < \frac{L}{2}, t \right) + \alpha_2 \left( -\frac{L}{2} < x < \frac{L}{2}, t \right) \nn\\
&& - i \sqrt{\frac{\gamma}{2 v_g}} \left[ e \left( t - \left[x + \frac{L}{2} \right] / v_g \right) + e \left( t - \abs{x-\frac{L}{2}} / v_g \right) \right]. \quad\quad
\eea
The total energy stored as SAW phonons between the two legs is
\bea
E_P &=& \hbar \omega_0 \int_{-L/2}^{L/2} \abssq{\alpha(x, t)} dx \nn\\
&=& \hbar \omega_0 \int_0^L \abssq{\alpha(x - L/2, t)} dx \nn\\
&=& \frac{\gamma \hbar \omega_0}{2 v_g} \int_0^L \abssq{e \left( t - x / v_g \right) + e \left( t - \abs{x - L} / v_g \right)} dx \nn\\
&=& \frac{\gamma \hbar \omega_0}{2 v_g} \left[ \int_0^L \abssq{e(t - x / v_g)} dx + \int_0^L \abssq{e \left(t - \abs{x - L} / v_g \right)} dx \right] \nn\\
&& + \frac{\gamma \hbar \omega_0}{v_g} \mathrm{Re} \left[ \int_0^L e(t - x / v_g) e^*(t - \abs{x-L} / v_g) dx \right].
\eea
In parameter region $D$, two SAW wave-packets are generated from the two legs of giant atom. When the two wave packets are well separated, we can neglect the overlap integral and have
\bea
E_P(t) &\approx& \frac{\gamma \hbar \omega_0}{2 v_g} \left[ \int_0^L \abssq{e(t - x / v_g)} dx \right. \nn\\
&& \left. + \int_0^L \abssq{e(t - \abs{x-L}/v_g)} dx \right] \nn\\
&\approx& \gamma \hbar \omega_0 \int_0^T \abssq{e(t - \tau)} d\tau.
\eea
For $m \leq t/T<m+1$, the time evolution of giant atom is $e(t) = \sum_{m = 0} e_m(t)$ with $e_m(t)$ given by \eqref{bm}. We neglect the overlap between different $e_m(t)$. The stored SAW energy at time $t = (m+1) T$ can then be calculated as
\bea
E_P &\approx& \gamma \hbar \omega_0 \int_0^T \abssq{e(t - \tau)} d\tau \nn\\
&\approx& \gamma \hbar \omega_0 \abssq{e(0)} \int_{m T}^{(m + 1) T} \abssq{e_m(t)} dt \nn\\
&\approx& \gamma \hbar \omega_0 \abssq{e(0)} \frac{\left( \gamma T e^{\gamma T} \right)^{2m}}{(m!)^2} \int_{m T}^{(m + 1) T} e^{-2 \gamma T \frac{t}{T}} \left(m - \frac{t}{T} \right)^{2m} dt \nn\\
&\approx& \gamma \hbar \omega_0 \abssq{e(0)} T \frac{\left( \gamma T e^{\gamma T} \right)^{2m}}{(m!)^2} \frac{(2m)!}{(-2 \gamma T)^{2m}} \int_{m}^{m + 1} e^{-2 \gamma T x} dx \nn\\
&\approx& \gamma \hbar \omega_0 \abssq{e(0)} T \frac{\left( \gamma T e^{\gamma T} \right)^{2m}}{(m!)^2} \frac{(2m)!}{(-2 \gamma T)^{2m}} \frac{1}{2 \gamma T} e^{-2 m \gamma T} \nn\\
&\approx& \frac{\gamma \hbar \omega_0 \abssq{e(0)}}{2 \gamma} \frac{\left( \gamma T e^{\gamma T} \right)^{2m}}{2\pi m (m/e)^{2m}} \frac{\sqrt{4 \pi m}(2m/e)^{2m}}{(-2 \gamma T)^{2m}} e^{-2 m \gamma T} \nn\\
&\approx& \frac{\hbar \omega_0 \abssq{e(0)}}{2 \sqrt{\pi}} \frac{1}{\sqrt{m}} \approx \frac{\hbar \omega_0 \abssq{e(0)}}{2 \sqrt{\pi}} \left(\frac{t}{T} - 1 \right)^{-1/2} \nn\\
&\approx& \frac{\hbar \omega_0 \abssq{e(0)}}{2 \sqrt{\pi}} \left(\frac{t}{T} \right)^{-1/2}.
\eea
Here, we have used Stirling's formula $m! = \sqrt{2 \pi m} (m/e)^m$ and the fact that $\gamma T \gg 1$ in parameter region $D$. Therefore, the stored energy follows a universal polynomial decay law $\propto t^{-1/2}$.

In parameter region $D$, as shown in Fig.~\ref{figSpectrum}(c1), the giant atom exhibits revival behaviour. We find that the revival peaks also decay polynomially. From Eq.~(\ref{bm}), the probability of the giant atom to be in the excited state in the time interval $mT \leq t < (m+1) T$, with $m\in\Z^+$, is
\be
P_e(t)\equiv |e_m(t)|^2 \approx  e^{-2(\gamma T) \frac{t}{T}} \frac{(\gamma T e^{\gamma T})^{2m}}{(m!)^2} \left(m - \frac{t}{T} \right)^{2m}.
\label{}
\ee
Here, we have assumed the giant atom is in the excited state initially, i.e., $e(0)=1$. From Eq.~(A43), we see that the atom follows the following general behaviour in each time interval: it starts in the ground state [$P_e(mT) = 0$], revives to a peak value $P_e^{max}$, and then decays exponentially at a rate $2\gamma$. The peak's position in the time interval $mT \leq t < (m+1) T$ is readily found to be
\begin{eqnarray}\label{tmax}
\frac{t_{m}}{T}=m+\frac{m}{\gamma T}.
\end{eqnarray}
The peak value $P_{e}^{max}(t_m)$ is given by
\begin{eqnarray}\label{polynomialDecaype}
P_{e}^{max}(t_{m})\approx
\Big(\frac{m^{m}}{e^mm!}\Big)^2\approx \frac{1}{2\pi m}=(1+\frac{1}{\gamma T})\frac{T}{2\pi t_{m}}.
\end{eqnarray}
Here, we have used $m!\approx \sqrt{2\pi m}(m/e)^m$ for $m\gg 1$.
We see that $P_{e}^{max}(t_m)$ follows a polynomial decay law $\propto t_{m}^{-1}$. We also see from Eq.(\ref{tmax}) that the time of the peak shifts
with $m$. When $m\approx \gamma T$, the peak is at $(m+1)T$, the boundary of the interval, which indicates that the decay behaviour changes around this value of $m$.

In the experiment, one can measure the outgoing phonons from the two legs of the giant atom. The outgoing phonon field for $m\leq t/T<m+1$ ($m\geq 1$) is given by
\begin{eqnarray}\label{}
\alpha^{out}(t)=-i\sqrt{\frac{\gamma}{2v_g}}\Big[e_m(t)+e_{m-1}(t-T)\Big].
\end{eqnarray}
Using Eq.~(\ref{bm}), we have
\begin{eqnarray}\label{}
\alpha^{out}(t)&\approx& -i\sqrt{\frac{\gamma}{2v_g}}
e_m(t)\Big[1+\frac{m}{\gamma T(m-t/T)}\Big].
\end{eqnarray}
We calculate energy accumulation of outgoing phonons during the time $m\leq
t/T<m+1$:
\begin{eqnarray}\label{Eoutpolynomial}
\frac{E^{out}(m)}{\hbar\omega_0}&\equiv&v_g\int_{mT}^{(m+1)T}|\alpha^{out}(t)|^2dt\nonumber\\
&=&\frac{\gamma}{2}\Big[\int_{mT}^{(m+1)T}|e_m(t)|^2dt+\int_{(m-1)T}^{mT}|e_{m-1}(t)|^2dt\nonumber\\
&&+\int_{mT}^{(m+1)T}\Big(e_m(t)e^*_{m-1}(t)+e^*_m(t)e_{m-1}(t)\Big)dt\Big]\nonumber\\
&=&\frac{\gamma}{2}\Big[\frac{1}{2\gamma\sqrt{\pi}
}\frac{1}{\sqrt{ m}}+\frac{1}{2\gamma\sqrt{\pi}
}\frac{1}{\sqrt{ m-1}}\nonumber\\
&& +\int_{mT}^{(m+1)T}|e_m(t)|^2\frac{2m}{\gamma T(m-t/T)}dt\Big]
\nonumber\\
&=&\frac{\gamma}{2}\Big[\frac{1}{2\gamma\sqrt{\pi}
}\frac{1}{\sqrt{ m}}+\frac{1}{2\gamma\sqrt{\pi}
}\frac{1}{\sqrt{ m-1}}\nonumber\\
&& +\frac{2m(\gamma Te^{\gamma T})^{2m}}{\gamma
T(m!)^2}\int_{mT}^{(m+1)T} e^{-2(\gamma
T)\frac{t}{T}}(m-\frac{t}{T})^{2m-1}dt\Big]\nonumber\\
&\approx&\frac{\gamma}{2}\Big[\frac{1}{2\gamma\sqrt{\pi}
}\frac{1}{\sqrt{ m}}+\frac{1}{2\gamma\sqrt{\pi}
}\frac{1}{\sqrt{ m-1}}\nonumber\\
&& +\frac{2m(\gamma Te^{\gamma T})^{2m}}{\gamma
T(m!)^2}T\frac{(2m-1)!}{(-2\gamma
T)^{2m-1}}\frac{e^{-2m\gamma T}}{2\gamma T}\Big]\nonumber\\
&=&\frac{\gamma}{2}\Big[\frac{1}{2\gamma\sqrt{\pi}
}\frac{1}{\sqrt{ m-1}} -\frac{1}{2\gamma\sqrt{\pi}
}\frac{1}{\sqrt{ m}} \Big]\nonumber\\
&\approx&\frac{1}{8\sqrt{\pi}}m^{-\frac{3}{2}}.
\end{eqnarray}
The above result shows that the energy of the outgoing phonons also follows a polynomial decay law $\propto m^{-\frac{3}{2}}$.

\subsection{Reflectance and transmittance}
\label{app:SinglePhononRTCoefficients}

To study the scattering properties of the giant atom, we send a right-propagating SAW towards the left leg, \textit{i.e.}, $\alpha_{1 \omega}(0) \neq 0$ and $\alpha_{2 \omega}(0) = 0$. Assuming the giant atom to be in the ground state initially, we have from \eqref{appet} that its dynamics are given by
\be
e(t) = 2 \sqrt{\frac{\gamma}{4\pi}} \int d\omega \frac{\alpha_{1 \omega}(0) \cos (\omega T/2)}{\omega - \omega_0 + i \gamma + i \gamma e^{i \omega T}} e^{-i \omega t}.
\ee
From \eqref{sawfield} we obtain the dynamics of forward-scattered SAWs
\bea
&&{\alpha}_{1\omega}(t) = e^{-i \omega t} \alpha_{1 \omega}(0) - i \frac{\gamma}{\pi} \cos \left( \frac{\omega T}{2} \right) \times \nn\\
&& \int d\omega' \frac{\alpha_{1 \omega'}(0) \cos \left( \frac{\omega' T}{2} \right)}{\omega' - \omega_0 + i \gamma + i \gamma e^{i \omega' T}} e^{-i \omega t} \int_0^t dt' e^{i (\omega - \omega') t'},
\eea
and the backward-scattered SAWs
\bea
&& \alpha_{2 \omega}(t) = - i \frac{\gamma}{\pi} \cos \left( \frac{\omega T}{2} \right) \times \nn\\
&& \int d\omega' \frac{\alpha_{1 \omega'}(0) \cos \left( \frac{\omega' T}{2} \right)}{\omega' - \omega_0 + i \gamma + i \gamma e^{i \omega' T}} e^{-i \omega t} \int_0^t dt' e^{i (\omega - \omega') t'}. \quad
\eea
In the long-time limit, we have
\bea
\alpha_{1 \omega}(\infty) &=& e^{-i \omega t} \alpha_{1 \omega}(0) \left( 1 - i \gamma \frac{1 + \cos \omega T}{\omega - \omega_0 + i \gamma + i \gamma e^{i \omega T}} \right), \qquad
\label{alpha1t} \\
\alpha_{2 \omega}(\infty) &=& e^{-i \omega t} \alpha_{1 \omega}(0) \left(-i \gamma \frac{1 + \cos \omega T}{\omega - \omega_0 + i \gamma + i \gamma e^{i \omega T}} \right). \qquad
\label{alpha2t}
\eea
We define the transmittance and reflectance as $\mathcal{T} \equiv \frac{\abssq{\alpha_{1 \omega}(+\infty)}}{\abssq{\alpha_{1 \omega}(0)}}$ and $\mathcal{R} \equiv \frac{\abssq{\alpha_{2 \omega}(+\infty)}}{\abssq{\alpha_{1 \omega}(0)}}$, respectively. They can be calculated from Eqs.~(\ref{alpha1t}) and (\ref{alpha2t}), giving
\bea
\mathcal{T} &=& \frac{(\omega_d - \omega_0 - \gamma \sin \omega_d T)^2}{(\omega_d - \omega_0 - \gamma \sin \omega_d T)^2 + \gamma^2 (1 + \cos \omega_d T)^2}, \quad \\
\mathcal{R} &=& \frac{\gamma^2 (1 + \cos \omega_d T)^2}{(\omega_d - \omega_0 - \gamma \sin \omega_d T)^2 + \gamma^2 (1 + \cos \omega_d T)^2}. \quad
\eea
It is easily seen that $\mathcal{T} + \mathcal{R} = 1$.

\section{Input-output theory for a giant atom}
\label{app:inout}

%\alg{[Maybe some of this should be in Sect. II?]}

In this appendix we give some further details on the input-output theory for the giant atom with arbitrary driving strength. We write the model atom-phonon Hamiltonian from \eqref{Hstart} in the main text as
\be
\begin{aligned}
& \mathscr{H} = H_S + \sum_{\alpha = 1,2} \int_0^\infty \dd\omega \hbar \omega a_{\alpha \omega}^\dag a_{\alpha \omega} \\
& + \hbar \sum_{\alpha = 1,2} \int_{-\infty}^\infty \dd\omega \sqrt{\frac{\gamma}{4\pi}} \left[ \sm a_{\alpha \omega}^\dag \left( e^{-i  \omega T/2} + e^{i  \omega T/2} \right) + \text{H.c.} \right],
\end{aligned}
\label{eq:inout:H}
\ee
where $H_S = \hbar \omega_0 \ketbra{e}{e}$ is the bare atom Hamiltonian. In \eqref{eq:inout:H}, we have made the standard Markov approximations of taking the atom's decay rate to be frequency independent and extended the lower integration limit to minus infinity for the interaction term \cite{Collett1984}.

How the phonon transmission line serves as a feedback loop for the atom is apparent when the equations of motion are formulated using the usual quantum optics input-output formalism. Following the standard approach \cite{Collett1984}, we find from the Hamiltonian in \eqref{eq:inout:H} that the quantum Langevin equation for an aribtrary atomic operator $x$ is
\be
  \begin{aligned}
    \dot{x}(t) = {} & \frac{i}{\hbar} [H_S, x] \\
    &+ \sum_{i=A,B} \Bigg\{ [\sp, x]
      \bigg[ \frac{\gamma}{2} \left(\sm(t) + \sm(t - T) \right) \\
      & \quad + i \sqrt{\frac{\gamma}{2}} \left( a_i^{\rm in}(t) + a_i^{\rm in}(t - T) \right) \bigg] \\
      & - \bigg[ \frac{\gamma}{2} \left(\sp(t) + \sp(t - T) \right) \\
      & \quad - i \sqrt{\frac{\gamma}{2}} \left( a_i^{{\rm in}\dag}(t) + a_i^{{\rm in}\dag}(t - T) \right) \bigg]
    [\sm, x] \Bigg\},
  \end{aligned}
\label{eq:Langevin}
\ee
where we have defined input free phonon fields incident on the atom at leg $A$ and leg $B$
\bea
  a_A^{\rm in}(t) \equiv {} & \frac{1}{\sqrt{2\pi}} \int_{-\infty}^\infty \dd\omega e^{-i \omega (t - t_0)} a_{1\omega}(t_0), \\
  a_B^{\rm in}(t) \equiv {} & \frac{1}{\sqrt{2\pi}} \int_{-\infty}^\infty \dd\omega e^{-i \omega (t + T - t_0)} a_{2\omega}(t_0),
\eea
where $t_0$ is some early time where the Heisenberg picture and Schr\"odinger picture operators coincide (we assume $t_0 < t - T$ for all $t$). We can similarly define output fields at leg $A$ and leg $B$ of the giant atom
\bea
  a_A^{\rm out}(t) \equiv
   {} & \frac{1}{\sqrt{2\pi}} \int_{-\infty}^\infty \dd\omega e^{-i \omega(t - t_1)} a_{2\omega}(t_1),\\
  a_B^{\rm out}(t) \equiv
   {} & \frac{1}{\sqrt{2\pi}} \int_{-\infty}^\infty \dd\omega e^{-i \omega(t - T - t_1)} a_{1\omega}(t_1),
\eea
where $t_1 > t + T$ is some late time. The output fields are given by the inputs and the atomic dynamics through input-output equations
\bea
  a_A^{\rm out}(t) = {} & a_B^{\rm in}(t - T) - i \sqrt{\frac{\gamma}{2}} \left[ \sm(t) + \sm(t - \tau) \right], \\
  a_B^{\rm out}(t) = {} & a_A^{\rm in}(t - T) - i \sqrt{\frac{\gamma}{2}} \left[ \sm(t) + \sm(t - \tau) \right] .
\eea
These boundary conditions are similar to the boundary conditions (\ref{connection3a}) and (\ref{connection3}) but replacing the single phonon excitation amplitude $e(t)$ by the lowering operator $\sm(t)$.

Equation (\ref{eq:Langevin}) is a non-linear quantum Langevin delay differential equation, making the feedback mechanism of the transmission line quite clear. It is, however, not easily solved in general. Note that in the presence of coherent input drives, these can conveniently be moved into the system Hamiltonian
\be
  \begin{aligned}
  H_S = {} & \hbar \omega_0 \ketbra{e}{e} \\
  & + \hbar V \sum_\alpha \left[ \left( \alpha_\alpha^{\rm in}(t)^*
  + e^{-i \varphi} \alpha_\alpha^{\rm in}(t - T)^* \right) \sm + \text{H.c.} \right],
  \end{aligned}
\ee
by displacing the phonon fields $a_{\alpha \omega} \to a_{\alpha \omega} + \sqrt{4\pi/\gamma} V \alpha_\alpha^{\rm in}(t)\delta(\omega - \omega_{d_\alpha})$ with $\alpha_1^{\rm in}(t) = A_1 e^{-i \omega_{d_1} t}$ and $\alpha_2^{\rm in}(t) = A_2 e^{-i \omega_{d_2}(t + T)}$.

In \secref{sect:cascade} we also used the fact that the problem can be mapped onto a setup with only a single feedback loop. This is easily seen from \eqref{eq:inout:H} by defining a new phonon operator ${b}_\omega \equiv \left(a_{1\omega} + a_{2\omega} e^{-i \omega T} \right)/\sqrt{2}$. In terms of this new field, the quantum Langevin equation can be written
\be
  \begin{aligned}
    \dot{x}(t) = {} & \frac{i}{\hbar}[H_S, x] \\
    & + \Bigg\{ [\sp, x]
      \bigg[ \gamma \left(\sm(t) + \sm(t - T) \right) \\
      & \quad + i \sqrt{\gamma} \left( b^{\rm in}(t) + b^{\rm in}(t - T) \right) \bigg] \\
      & - \bigg[ \gamma \left(\sp(t) + \sp(t - T) \right) \\
      & \quad - i \sqrt{\gamma} \left( b^{{\rm in} \dag}(t) + b^{{\rm in}\dag}(t - T) \right) \bigg]
    [\sm, x] \Bigg\},
  \end{aligned}
  \label{eq:Langevin2}
\ee
where
\be
    b^{\rm in}(t) \equiv \frac{1}{\sqrt{2\pi}} \int_{-\infty}^\infty \dd\omega e^{-i \omega(t - t_0)} b_{\omega}(t_0)
    = \frac{1}{\sqrt{2}} \left[ a_A^{\rm in}(t) + a_B^{\rm in}(t) \right].
\ee
We can also define an output field given by the usual input-output equation
\be
  \begin{aligned}
    b^{\rm out}(t) = {} & b^{\rm in}(t - T) - i \sqrt{\gamma}\left[ \sm(t) + \sm(t - \tau) \right] \\
    = {} & \frac{1}{\sqrt{2}} \left[ a_A^{\rm out}(t) + a_B^{\rm out}(t) \right].
  \end{aligned}
\ee
Note that in a situation where the left and right input fields are identical, we have, based on symmetry, that $a_A^{\rm out}(t) = a_B^{\rm out}(t) = b^{\rm out}(t)/\sqrt{2}$.

\section{Computing output-field correlation functions from the cascaded master equation}
\label{app:outfield}

In this appendix, we outline how output field correlation functions can be computed from \eqref{eq:cascade} building on the method presented in Ref.~\cite{Grimsmo2015}. We want to calculate an output field correlation function of the type
\be
  \expec{c_1(t_1) c_2(t_2) \dots c_n(t_n)},
  \label{eq:outfieldcorr}
\ee
where $c_i(t_i)$ is one of $b^{\rm out}(t_i)$ or $b^{{\rm out}}(t_i)^\dag$ (see \appref{app:inout}) and we assume that all the times are different, $t_i \neq t_j$, such that the time-ordering is arbitrary (equal times can be taken as a limit). First, it is illustrative to recall how this can be done for a conventional Markovian open quantum system where the evolution is given by a Lindblad master equation
\be
  \frac{\dd}{\dd t} \mathcal{E}(t) = \left\{ -\frac{i}{\hbar} \mathcal{H}[H_s(t)] + \mathcal{D}[{L}] \right\} \mathcal{E}(t),
\ee
for the time-propagator $\mathcal{E}(t)$ [\textit{i.e.}, the state at time $t$ is given by $\rho(t) = \mathcal{E}(t) \rho(0)$] and where the output field is given by the input field and the system dynamics through an input-output equation
\be
  b^{\rm out}(t) = b^{\rm in}(t) + L(t).
\ee
As is well known, \eqref{eq:outfieldcorr} can be computed through the so-called quantum regression formula \cite{Collett1984, Gardiner2004}
\be
  \begin{aligned}
  & \expec{c_1(t_1) c_2(t_2) \dots c_n(t_n)} \\
  & = \expec{\mathcal{F}_n \mathcal{E}(s_n - s_{n - 1}) \dots \mathcal{F}_2 \mathcal{E}(s_2 - s_1) \mathcal{F}_1 \mathcal{E}(s_1) \rho(0)},
  \end{aligned}
\ee
where we have time-ordered the times and relabelled them by $s_1 < s_2 < \dots < s_n$, and defined
\bea
  \mathcal{F}_j \rho = {} & L \rho \qquad \text{for } c_j(s_j) = b^{\rm out}(s_j), \\
  \mathcal{F}_j \rho = {} & \rho L^\dagger \qquad \text{for } c_i(s_j) = b^{\rm out}(s_j)^\dag.
\eea

The computation for a system with time-delay, with the time-propagator master equation \eqref{eq:cascade}, is entirely analogous. First, we define new time-variables $t_i^*$ through
\be
  t_i^* = t - l_i T,
\ee
where $l_i = \left[\frac{t}{T}\right]$ is the largest integer less than or equal to $\frac{t}{T}$, and time-order these auxiliary time-variables from earliest to latest and call them $s_1^* < s_2^* < \dots < s_n^*$. The correlation function in \eqref{eq:outfieldcorr} is then given by
\be
  \begin{aligned}
  & \expec{c_1(t_1) c_2(t_2) \dots c_n(t_n)} \\
  & = \expec{\mathcal{F}_n \mathcal{E}_T(s_n^* - s_{n - 1}^*) \dots \mathcal{F}_2 \mathcal{E}_T(s_2^* - s_1^*) \mathcal{F}_1 \mathcal{E}_T(s_1^*) \rho(0)},
  \end{aligned}
\label{eq:delayedregression}
\ee
where now
\bea
  \mathcal{F}_j\rho = {} & L_{l_{j-1},l_j} \rho \qquad \text{for } c_j(t_j) = b^{\rm out}(t_j), \\
  \mathcal{F}_j\rho = {} & \rho L_{l_{j-1},l_j}^\dag \qquad \text{for } c_j(t_j) = b^{\rm out}(t_j)^\dag,
\eea
and the index $j$ refers to the time-ordering of $s_j^*$. A formal proof of \eqref{eq:delayedregression} can be given using the tensor-network representation of the time-propagator used in Ref.~\cite{Grimsmo2015} and will be presented in a future work \cite{Whalen2016}.

\bibstyle{apsrev4-1}
%\bibliography{GiantAtomRefs}

%

\end{document}